\documentclass[aps,pra,twocolumn,floatfix,superscriptaddress]{revtex4-1}
\usepackage{amsfonts}
\usepackage{graphicx}
\usepackage{amsmath}
\usepackage{amssymb}
\usepackage{bm}
\usepackage{amsthm}
\usepackage{color}
\usepackage[dvipsnames]{xcolor}
\usepackage{mathrsfs,latexsym}
\usepackage{soul}
\usepackage{verbatim}
\usepackage{revsymb}
\usepackage{mhchem}
\usepackage{hyperref}

\hypersetup{
    colorlinks=true,           linkcolor=blue,              citecolor=red,            urlcolor=blue        }

\usepackage{ulem}

\begin{document}

\title{Collective quantum phase slips in multiple nanowire junctions}
\author{Zeng-Zhao Li}
\affiliation{Division of Solid State Physics and NanoLund, Lund University, Box 118, S-22100 Lund, Sweden} 
\affiliation{Interdisciplinary Center of Quantum Information and Zhejiang Province Key Laboratory of Quantum Technology and Device, Department of Physics and State Key Laboratory of Modern Optical Instrumentation, Zhejiang University, Hangzhou 310027, China}
\author{Tie-Fu Li}
\affiliation{Institute of Microelectronics and Tsinghua National Laboratory of Information Science and Technology, Tsinghua University, Beijing 100084, China}
\author{Chi-Hang Lam}
\affiliation{Department of Applied Physics, Hong Kong Polytechnic University, Hung Hom, Hong Kong, China}
\author{J. Q. You}
\email{jqyou@zju.edu.cn}
\affiliation{Interdisciplinary Center of Quantum Information and Zhejiang Province Key Laboratory of Quantum Technology and Device, Department of Physics and State Key Laboratory of Modern Optical Instrumentation, Zhejiang University, Hangzhou 310027, China}
\date{\today }

\begin{abstract}
{Realization of robust coherent quantum phase slips represents a significant experimental challenge. Here we propose a new design consisting of multiple nanowire junctions to realize a phase-slip flux qubit. It admits good tunability provided by gate voltages applied on superconducting islands  separating nanowire junctions. In addition, the gates and junctions can be identical or distinct to each other leading to symmetric and asymmetric setups.
We find that the asymmetry can improve the performance of the proposed device, compared with the symmetric case. In particular, it can enhance the effective rate of collective quantum phase slips. Furthermore, we demonstrate how to couple two such devices via a mutual inductance. This is potentially useful for quantum gate operations. Our investigation on how symmetry in multiple nanowire junctions affects the device performance should be useful for the application of phase-slip flux qubits in quantum information processing and quantum metrology.}
\end{abstract}

\maketitle


\section{Introduction}

There has been a growing research interest on  quantum phase slips in not only condensed matters but also ultracold quantum gas \cite{DErrico17PhilTransRSocA}. In the solid-state context such as  superconducting nanowires, the phase of the superconducting order parameter $\psi = {\psi _0}{e^{i\phi }}$ is allowed to change (i.e., slip) rapidly by $\pm 2\pi$ if its amplitude tends to zero, due to the requirement that $\psi _0^2\nabla \phi$ gives a constant \cite{Tinkham96,Bezryadin13,ArutyunovGolubevZaikin08PhysRep}. 
For a long time, achieving coherent quantum phase slips has been a  challenging topic. Traditional methods~\cite{NewbowerTinkham1972PRB} rely on the detection of phase-slip changes in resistance measurements of superconducting wires. However, one cannot fully reveal the quantum nature of the phase-slip process~\cite{Lau01PRL,Bollinger08PRL} because phase slips can also be activated by thermodynamic fluctuations that contribute to the residual resistance of superconducting wires~\cite{LangerAmbegaokar67PR,McCumberHalperin70PRB,SkocpolBeasleyTinkham74JLowTempPhys}. 
A more sophisticated method for detecting coherent quantum phase slips is to engineer a device known as a quantum phase-slip junction~\cite{MooijHarmans05NJP}. This phase-slip junction can play the role of a Josephson junction~\cite{Likharev86} in a superconducting flux qubit to form a new kind of qubit known as the phase-slip flux qubit~\cite{MooijHarmans05NJP,MooijNazarov06Nphys}. Compared with conventional superconducting qubits for quantum information processing (see Refs.~\onlinecite{YouNori11nature,ClarkeWilhelm08nature,DevoretSchoelkopf13science} for reviews), this phase-slip qubit is insensitive to the charge noise.
Moreover, similar to Josephson junctions for accurate standards of voltage~\cite{KautzLloyd1987APL}, the flux-charge duality~\cite{MooijNazarov06Nphys,Kerman2013NJP} renders this phase-slip qubit a promising device for providing a quantum current standard~\cite{Zimmerman2005PhysTod}. In recent experiments, highly disordered indium oxide ($\ce{InO_{x}}$)
and nibium nitride ($\ce{NbN}$) nanowires were utilized to achieve  phase-slip flux qubits~\cite{AstafievTsai12nature,Bezryadin12nature,Peltonen13prb,Peltonen16PRB}. 
In addition, a single-charge transistor based on quantum phase slips was realized~\cite{HongistoZorin12PRL}, which is dual to dc SQUID and  can be operated as an electrometer. 
Quantum paired phase slips that could reduce decoherence in qubits by exploiting parity effects was experimentally demonstrated~\cite{BelkinBezryadin15PRX}. 
In particular, a very recent experiment, which used microscopic spectroscopy~\cite{AstafievTsai12nature,Peltonen13prb,Peltonen16PRB} rather than direct current transport measurement~\cite{HongistoZorin12PRL,Webster13PRB,Kafanov13JAP}, reported the realization of a charge quantum interference device~\cite{GraafAstafiev18nphys} based on two phase-slip junctions \cite{Zhao13CPL}. These experiments represent important steps towards the applications of phase-slip circuits in quantum information processing and
quantum metrology. 
Note that coherent quantum phase slips could also be explored through the approximate self-duality of Josephson junction circuits, such as a Cooper-pair box~\cite{BellGershenson16PRL} and Josephson arrays~\cite{ErgulHaviland13NJP,PopGuichard10nphys,Haviland10nphys,Pop12PRB,Manucharyan12PRB,RastelliHekking13PRB,MarcoHekking15PRB,SvetogorovHekking18PRB}, which however have been shown a challenging route towards a phase-slip quantum current standard~\cite{Cedergren17PRL}. 

Achieving robust coherent quantum phase slips is still an experimental challenge. The major difficulty comes from quasiparticle dissipations in nanowires or vortex cores which make the phase-slip rate imperceptible. These dissipations can be suppressed in a highly disordered superconductor near the superconductor-insulator transition, where electrons are localized and quantum fluctuations of the order parameter are prominent. Despite the weak quantum fluctuations in bulk disordered superconductors, they can become significantly stronger in disordered nanowires where the localization length is comparable to the coherence length. 
It was shown that the phase-slip rate can be increased by raising the disorders in the superconducting nanowires~\cite{Zaikin97prl,GolubevZaikin01prb}. Further increases are however prohibited since too much disorder in the nanowires can yield strong Coulomb interactions and destroy the superconductivity~\cite{Finkelstein94PhysB}.
Alternatively, the phase-slip rate can be increased by using a longer nanowire~\cite{Lau01PRL,MooijHarmans05NJP}, but the enhancement is also limited because the quantum fluctuations needed for the emergence of the phase slips become weakened and even disappear for long nanowires,  e.g., the $\ce{MoGe}$ nanowire can be up to $\zeta\sim200$ nm long ~\cite{Lau01PRL,Bezryadin08JPCM} while still maintaining the needed quantum fluctuations. Therefore, an alternative method to enhance the phase-slip rate is strongly desired. 

In this paper, we investigate the effects of symmetric and asymmetric setups on collective quantum phase slips in multi-junction phase-slip flux qubits. 
In contrast to a single junction exhibiting only weak phase slips, multiple junctions are particularly important because they can collectively give rise to a large phase-slip rate demonstrating appreciable quantum phase slips. In such a phase-slip qubit, each superconducting island separated by two adjoining phase-slip junctions is biased by a gate voltage, so as to achieve a tunable phase-slip rate. 
Moreover, we propose to couple two multi-junction phase-slip flux qubits via the mutual inductance between them. These inductively coupled phase-slip flux qubit pair are dual to a charge qubit pair coupled via a mutual capacitance. Our proposed multi-junction device has distinct advantages over a single phase-slip junction or a charge quantum interference device based on two phase-slip junctions. This is because various symmetry configurations can give rise to drastically distinct results and may potentially be used for example to achieve a large effective phase-slip rate. This can widen the range of materials usable for superconducting quantum circuits.

\section{Multi-junction Phase-slip Flux Qubit \label{sec:superPS}}

\begin{figure}[t]
\centering
\begin{minipage}[c]{0.5\textwidth}
  \centering
  \includegraphics[width=0.9\columnwidth]{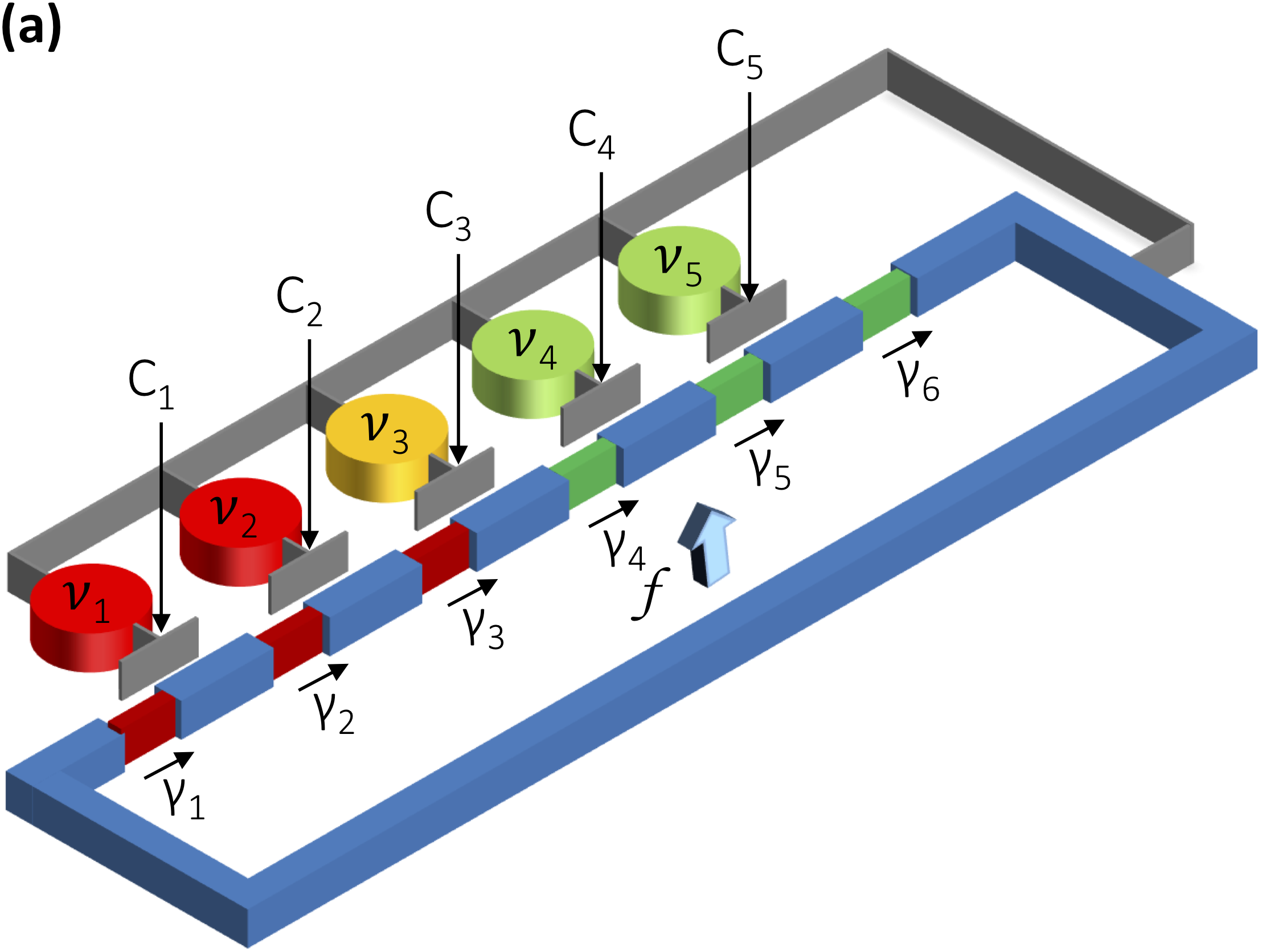}
\end{minipage}\newline
\begin{minipage}[c]{0.5\textwidth}
  \centering
  \includegraphics[width=0.9\columnwidth]{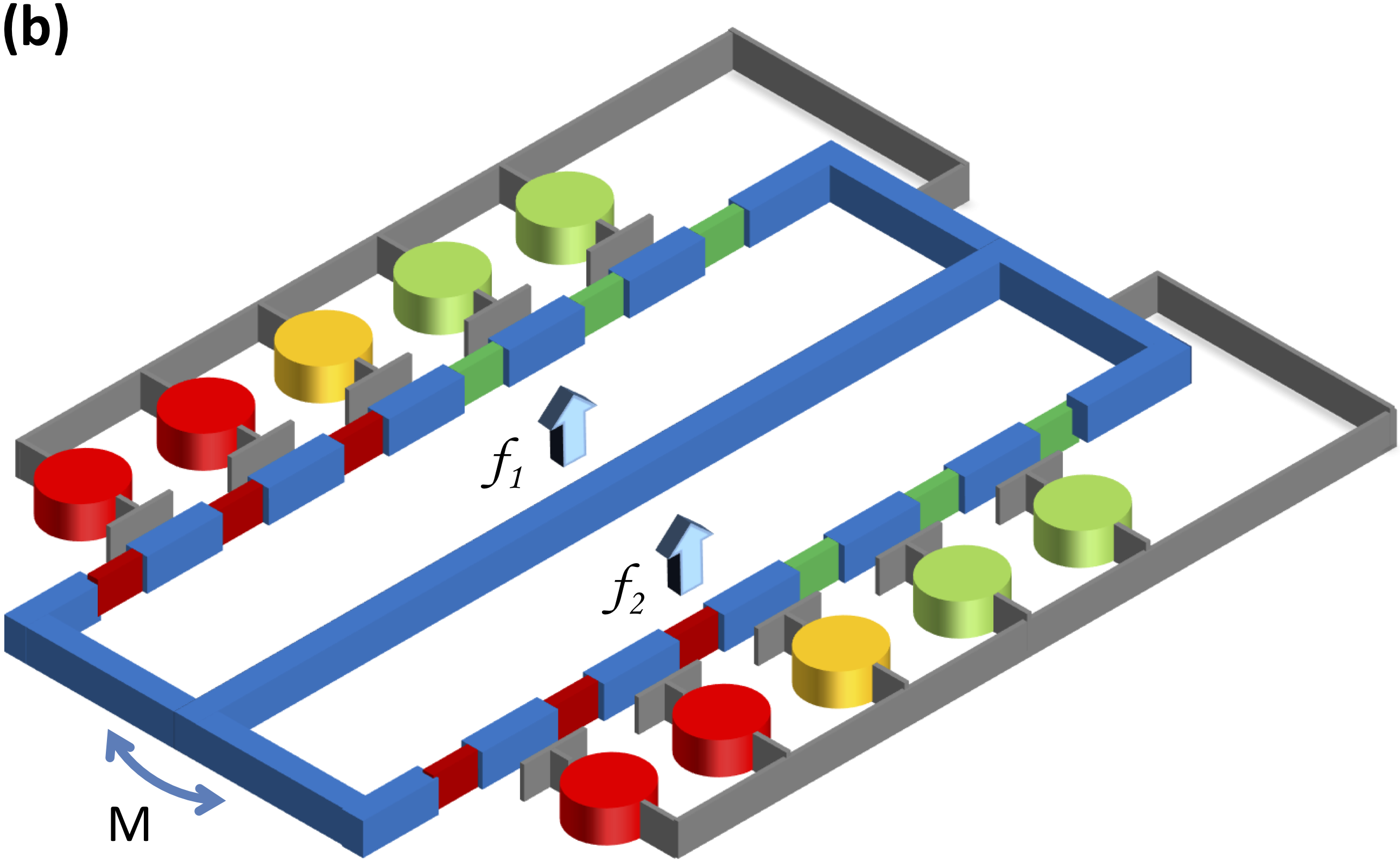}
\end{minipage}
\caption{(a) Schematic diagram of a multi-junction phase-slip flux qubit, where $\protect\gamma_{i}$, for $i=1,2,\ldots,m$ with $m=6$, represents the phase drop across the $i$th junction in the loop, and $f\equiv \Phi_{\mathrm{ext}}/\Phi_{0}$ is the reduced magnetic flux applied to the loop. Each superconducting island between two adjoining phase-slip junctions is controlled by the gate voltage $\nu_{l}$ via a gate capacitance $C_{l}$, where $l=1,2,\ldots,m-1$. (b) Two multi-junction phase-slip flux qubits coupled by a mutual inductance $M$ between them.}
\label{fig:QPSJ_fig}
\end{figure}

The proposed multi-junction phase-slip flux qubit is schematically shown in Fig.~\ref{fig:QPSJ_fig}(a), where a superconducting loop is interrupted by $m$ phase-slip junctions. The voltage drop across each junction is given by $V_{i}=V^C_{i}\sin \left( 2\pi q_{i}\right)$~\cite{MooijNazarov06Nphys}  for $i=1,2,\cdots ,m$, where $q_{i}$ is the number of Cooper pairs having tunnelled through the $i$th phase-slip junction.
Also, $V^C_{i}=2\pi E_{i}/2e$ is the critical voltage of the $i$th junction, where $E_{i}$ denotes the phase-slip rate. 
Neighboring phase-slip junctions are connected by a superconducting island biased by a gate voltage $\nu_l$ via a gate capacitance $C_{l}$ $\left( l=1,2,\ldots ,m-1\right) $. The reduced offset charge on each island is $N_{l}\equiv C_{l}\nu_{l}/2e$ and the supercurrent through each junction is $I=2e\dot{q}_i$. The phase drop across the $i$th phase-slip junction with a kinetic inductance $L_{ki}$ is given by $\gamma _{i}=2\pi \left( L_{ki}I/\Phi _{0}\right)$, and the phase drop related to the geometric inductance $L_{g}$ of the loop is $\gamma _{g}=2\pi \left( L_{g}I/\Phi _{0}\right) $, where $\Phi _{0}\equiv h/2e$ is the flux quantum.

We consider $m$ phase-slip junctions in which the first $\alpha$ of them ($i=1,2, \dots \alpha$) have properties  $\{\gamma_A,L_{kA},E_{A},V^C_{A}\}$, while the other $m-\alpha$ junctions ($i=\alpha+1,\alpha+2,\dots m$) are characterized by  $\{\gamma_B,L_{kB},E_{B},V^C_{B}\}$, i.e., 
\begin{eqnarray}
&&\{\gamma_i,L_{ki},E_{i},V^C_{i}\}  \nonumber\\
= &&\Biggr\{ \begin{array}{ll}
\{\gamma_A,L_{kA},E_{A},V^C_{A}\} &\text{if } 1\le i\le\alpha \\
\{\gamma_B,L_{kB},E_{B},V^C_{B}\} &\text{if } \alpha < i\le m .
\end{array}
\label{eq:QPSJ_AB}
\end{eqnarray}
The two sets are colored in red and green respectively in Fig.~\ref{fig:QPSJ_fig}. An asymmetric number of junctions in the two sets (i.e., $\alpha\neq m/2$) can significantly change the behaviors of the proposed device and in particular improve its performance, as demonstrated below. 

Accordingly, we have three sets of values for the capacitance $C_{l}$ and the gate voltage $\nu_{l}$ defined by
\begin{eqnarray}
\{C_{l},\nu_{l}\} &=& \Biggr\{ \begin{array}{lll}
\{C_{A},\nu_{A}\} & \text{ if } 1\le l\le \alpha-1 \\
\{C_{C},\nu_{C}\} & \text{ if } l =\alpha \\
\{C_{B},\nu_{B}\} & \text{ if } \alpha+1\le l\le m-1 
\end{array}
\label{eq:capacitanceVoltage}
\end{eqnarray}
and the corresponding gates are shaded in red, yellow and green in Fig.~\ref{fig:QPSJ_fig}. 
For time-independent applied voltages at the gates, charge balance implies 
\begin{eqnarray}
q_{i+1}-q_i &=& \Biggr\{ \begin{array}{lll}
N_{A} = \frac{C_{A}\nu_{A}}{2e} & \text{ if } 1\le i\le \alpha-1 \\
N_{C} = \frac{C_{C}\nu_{C}}{2e} & \text{ if } i =\alpha \\
N_{B} = \frac{C_{B}\nu_{B}}{2e} & \text{ if } \alpha+1\le i\le m-1 
\end{array}
\label{eq:chargeImblance}
\end{eqnarray}
Note that the superconducting island with a reduced off-set charge $N_{C}$ connects the two sets of phase-slip junctions and is particularly important as will be demonstrated below. 

Adopting the fluxoid (flux quanta) representation, 
the fluxoid states $\{\left\vert n\right\rangle \}$ are the eigenstates of $n\equiv \Phi /\Phi _{0}$.  Let $f\equiv \Phi _{\mathrm{ext}}/\Phi _{0}$ be the reduced externally-applied flux of the loop.
The Hamiltonian of the multi-junction
phase-slip flux qubit is (see Appendix \ref{sec:Append_qubit})
\begin{eqnarray}
H_q &=&\sum_n \Big[E_{L}\left( n-f\right) ^{2} |n\rangle\langle n| \notag \\
&&  -\frac{1}{2}E_{\mathrm{eff}} \left( e^{-i\varphi} \left\vert n+1\right\rangle \left\langle n\right\vert 
+ e^{i\varphi} \left\vert n\right\rangle \left\langle n+1\right\vert \right) \Big],
\label{eq:Hamiltonian_1}
\end{eqnarray}%
with the inductive energy scale $E_{L}$, the effective phase-slip rate $E_{\rm eff}$, and the phase $\varphi$ given respectively by 
\begin{eqnarray}
E_L &=& \frac{\Phi_0^2}{2[L_g+\alpha L_{kA} +(m-\alpha)L_{kB}]}, \label{eq:inductiveE} \\
E_{\rm eff} &=& \sqrt{\eta_A^2+\eta_B^2 +2\eta_A\eta_B\cos(4\pi\bar{q})}, \label{eq:effective_tunneling}\\ 
\varphi &=& \arctan\left[\frac{\eta_A - \eta_B}{\eta_A + \eta_B} \tan(2\pi\bar{q})\right], \label{eq:varphi}
\end{eqnarray}
where
\begin{eqnarray}
\eta_A &=& \frac{E_{A}\sin(\alpha\pi N_{A})}{\sin(\pi N_{A})}, \label{eq:eta_X} \\
\eta_B &=& \frac{E_{B}\sin[(m-\alpha)\pi N_{B}]}{\sin(\pi N_{B})}, \label{eq:eta_Y} \\
\bar{q} &=& \frac{1}{2}N_{C} +\frac{m-\alpha-1}{4} N_{B} +\frac{\alpha-1}{4} N_{A}. \label{eq:bar_q}
\end{eqnarray}
Note that the phase $\varphi$ vanishes if we consider e.g., $E_{A}=E_{B}$, $\alpha=m/2$, and $N_{A}=N_{B}$. Then, $H_q$ in  Eq.~(\ref{eq:Hamiltonian_1}) reduces to the same form of the Hamiltonian with a real phase-slip rate for a single-junction phase-slip flux qubit~\cite{MooijHarmans05NJP,MooijNazarov06Nphys}. 
The same reduction can alternatively be obtained by applying the transformation
\begin{eqnarray}
|\widetilde{n}+1\rangle &=& e^{-i\varphi (n+1)} |n+1\rangle, 
|\widetilde{n}\rangle = e^{-i\varphi n} |n\rangle
\label{eq:tansform}
\end{eqnarray}
and Eq.~(\ref{eq:Hamiltonian_1}) becomes real and is given by
\begin{eqnarray}
H_q &=&\sum_{\widetilde{n}} \Big[E_{L}\left(n -f\right) ^{2} |\widetilde{n}\rangle\langle \widetilde{n}| \notag \\
&&-\frac{1}{2}E_{\mathrm{eff}} \left( \left\vert \widetilde{n}+1\right\rangle \left\langle
\widetilde{n}\right\vert +\left\vert \widetilde{n}\right\rangle \left\langle \widetilde{n}+1\right\vert \right)\Big] .
\label{eq:Hamiltonian_2}
\end{eqnarray}
Note that the above transformation does not alter the inductive energy scale $E_{\rm L}$  and the phase-slip rate $E_{\rm eff}$ which are given in Eqs.~(\ref{eq:inductiveE}) and (\ref{eq:effective_tunneling}), respectively. 
In the proposed phase-slip flux qubit, the inductive energy proportional to $E_L$ depends parabolically on the applied flux $\Phi _{\mathrm{ext}}=f \Phi _{0}$ at each fluxoid number and the phase-slip rate $E_{\rm eff}$ couples states with adjacent fluxoid numbers and lifts the degeneracy at half integer values of $f$. In the considered flux regime satisfying $E_{L}\gg E_{\mathrm{eff}}$, this proposed multi-junction phase-slip flux qubit, described by Eq.~(\ref{eq:Hamiltonian_1}) with a real effective phase slip rate [i.e., Eq.~(\ref{eq:Hamiltonian_2})], is dual to a Cooper-pair box~\cite{MooijNazarov06Nphys}. 

Compared with the single-junction phase-slip flux qubit \cite{MooijHarmans05NJP}, Eq.~(\ref{eq:inductiveE}) shows that $E_{L}$ can be decreased by using multiple phase-slip junctions, so that the effect of the flux noise on the qubit is suppressed. Also, note that the effective phase-slip rate given by Eq.~(\ref{eq:effective_tunneling}) depends on the reduced offset charges (i.e., $N_{A}$, $N_{B}$, and $N_{C}$) and also the numbers of junctions (i.e., $\alpha$ and $m$). 
Therefore, in comparison with the single-junction phase-slip flux qubit~\cite{MooijHarmans05NJP,MooijNazarov06Nphys}, the proposed qubit consisting of multiple junctions can be tuned by not only the externally applied magnetic flux $\Phi_{\mathrm{ext}}$ in the loop (i.e., $f$) but also the gate voltage on each island. 
We finally mention that the use of multiple junctions for the purpose of enhancing collectively quantum phase slips in our work is different from the shunted large-capacitance Josephson junctions in a fluxonium that behave effectively like a superinductance and are used to reduce charge fluctuations~\cite{Manucharyan09science}. 

\section{Collective quantum phase slips in the phase-slip flux qubit}

To illustrate collective phase slips of the proposed multi-junction phase-slip flux qubit and in particular how they are influenced by the symmetry property of the qubit, in this section we first consider two limiting cases, i.e., $\cos(4\pi\bar{q})=\pm 1$ in Eq.~(\ref{eq:effective_tunneling}), and then we investigate general cases of both symmetric and asymmetric setups. 

\begin{figure}
\includegraphics[width=0.9\columnwidth]{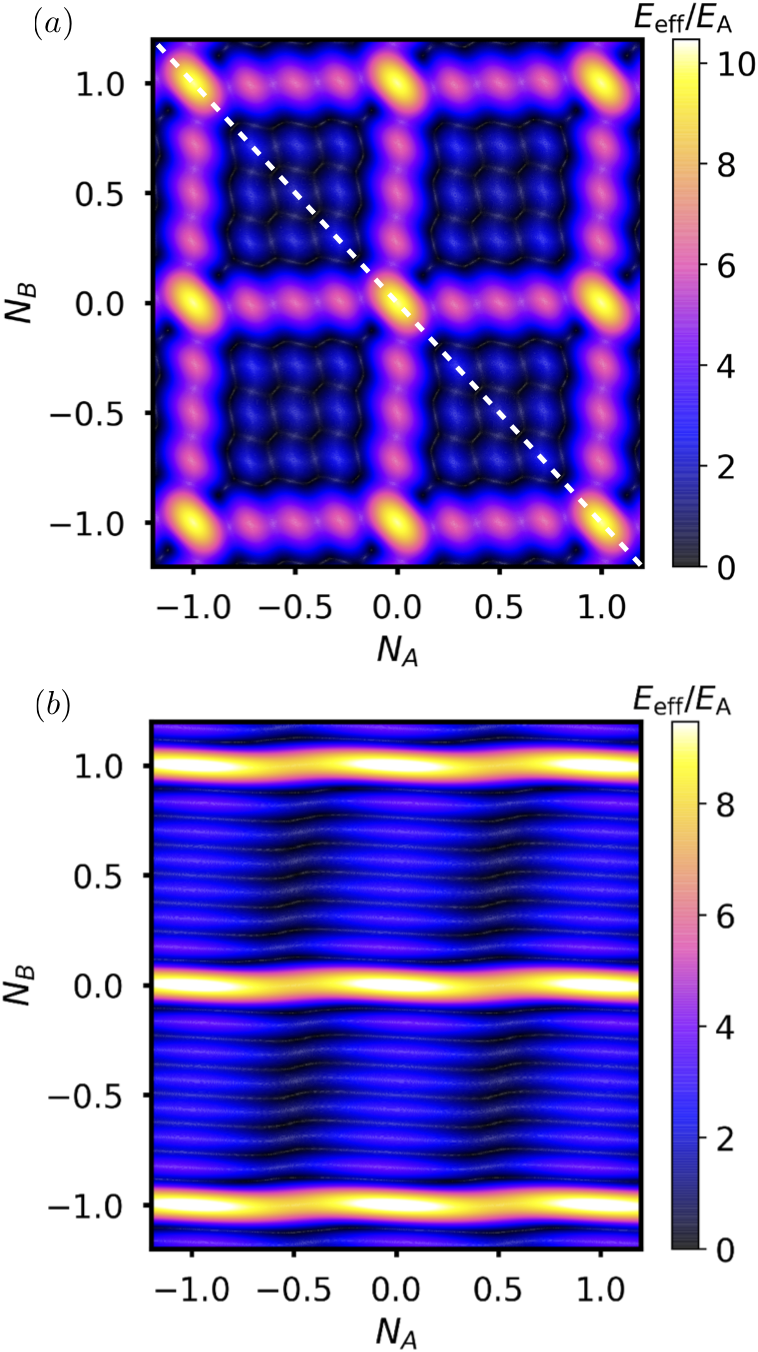}
\caption{The effective phase-slip rate $E_{\mathrm{eff}}$ of a multi-junction phase-slip flux qubit (in units of $E_{A}$) versus the reduced offset charges $N_{A}(\equiv C_{A}\nu_{A}/2e)$ and $N_{B}(\equiv C_{B}\nu_{B}/2e)$ for (a) $\alpha=5$ and (b) $\alpha=2$. Other parameters are $N_{C}=1$, $m=10$, and $E_{B}/E_{A}=1$.}
\label{fig:alpha5_2_EB1_symm}
\end{figure}

When considering the limit $\cos(4\pi\bar{q})=1$ that is accessible by tuning $N_C$ via the gate voltage $C_C$ [see Eq.~(\ref{eq:chargeImblance})] to be $N_C=k-(m-\alpha-1)N_B/2-(\alpha-1)N_A/2$ (with $k$ being an integer) obtained from Eq.~(\ref{eq:bar_q}), the effective phase-slip rate becomes $E_{\rm eff}=|\eta_A+\eta_B|$. The behavior of $\eta_A$ and $\eta_B$ with local maxima or minima in Eqs.~(\ref{eq:eta_X}) and (\ref{eq:eta_Y}) would give rise to periodic oscillations of $E_{\rm eff}$, as demonstrated below. If integer values of both $N_A$ and $N_B$ are considered, Eqs.~(\ref{eq:eta_X}) and (\ref{eq:eta_Y}) reduce to $\eta_A=\alpha E_A$ and $\eta_B=(m-\alpha)E_B$, respectively. We then have $E_{\rm eff}=\alpha E_A + (m-\alpha)E_B$, indicating constructively collective phase slips of multiple junctions. The further consideration of a symmetric setup, i.e., $E_A=E_B$ and $\alpha=m/2$ shows the $m$-fold enhancement [see for example the lattice points in Fig.~\ref{fig:alpha5_2_EB1_symm}(a)]. 

\begin{figure*}
\includegraphics[width=0.9\textwidth]{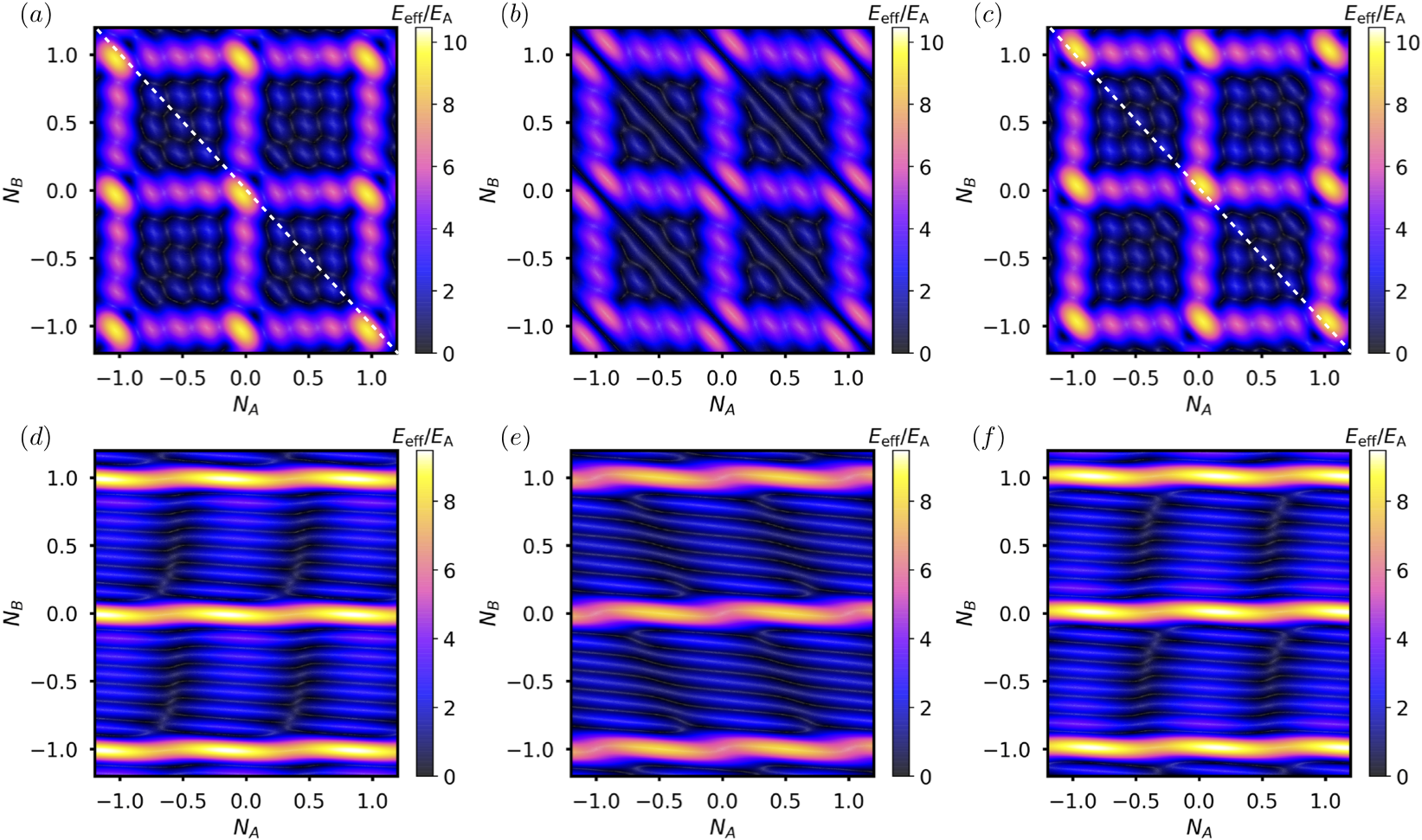}
\caption{The effective phase-slip rate $E_{\mathrm{eff}}$ of an asymmetric multi-junction phase-slip flux qubit (in units of $E_{A}$) versus the reduced offset charges $N_{A}(\equiv C_{A}\nu_{A}/2e)$ and $N_{B}(\equiv C_{B}\nu_{B}/2e)$ for (a,d) $N_{C}=0.2$, (b,e) $N_{C}=0.5$, (c,f) $N_{C}=0.8$ with $\alpha=5$ in (a,b,c) and $\alpha=2$ in (d,e,f). Other parameters are $m=10$ and $E_{B}/E_{A}=1$.}
\label{fig:alpha5_2_EB1_asymm}
\end{figure*}

For the other limit $\cos(4\pi\bar{q})=-1$ that is achievable via $N_C=k+1/2-(m-\alpha-1)N_B/2-(\alpha-1)N_A/2$, Eq.~(\ref{eq:effective_tunneling}) becomes $E_{\rm eff}=|\eta_A-\eta_B|$. For integer values of $N_A$ and $N_B$, it reduces to $E_{\rm eff}=|\alpha E_A - (m-\alpha)E_B|$, implying destructively collective phase slips of two sets of junctions defined in Eq.~(\ref{eq:QPSJ_AB}). This rate vanishes if a symmetric setup, i.e., $E_A=E_B$ and $\alpha=m/2$ ($m$ is an even integer) is considered [see, e.g., Fig.~\ref{fig:alpha5_2_EB1_asymm}(b)]. When $\alpha$ or $E_B$ goes away from the symmetric point $m/2$ or $E_A$ respectively, $E_{\rm eff}$ increases from zero to a finite value, as also shown in Figs.~\ref{fig:alpha_EB1}(a), \ref{fig:alpha_EB1}(b) or Fig.~\ref{fig:E_S_eff_vs_ESBA}(a), implying the advantage of an asymmetric setup. This asymmetry-induced enhancement of $E_{\rm eff}$ still  holds for $m$ being an odd number. 

In the following we present our detail study of collective phase slips by considering both symmetric and asymmetric cases. This would not only verify the analyses of limiting cases above but also illustrate how the symmetry property influences and particularly enhances the performance (i.e., collective coherent quantum phase slips) of the proposed phase-slip flux qubit. 

\subsection{Symmetric case \label{sec:symm}} 

The symmetric configuration of the device corresponds to an equal number in the two sets of phase-slip junctions (i.e., $\alpha=m/2$), identical single-junction phase-slip rate (i.e., $E_{A}=E_{B}$), and also an integer value for the reduced offset charge $N_{C}(=C_{C}\nu_{C}/2e)$. The first two conditions lead to a vanishing phase (i.e., $\varphi=0$) of the tunnelling flux in Eq.~(\ref{eq:Hamiltonian_1}) and identical coefficients of the $N_{A}$ and $N_{B}$ terms in the expression of $\bar{q}$ in Eq.~(\ref{eq:bar_q}). 

In Fig.~\ref{fig:alpha5_2_EB1_symm}(a), the effective phase-slip rate from Eq.~(\ref{eq:effective_tunneling}) for a symmetric setup is presented. It is shown that the tuning of both $N_{A}$ and $N_{B}$ generates square lattice patterns. These lattice patterns essentially result from the symmetric dependence of $E_{\rm eff}$ on $N_{A}$ and $N_{B}$. In particular, $E_{\rm eff}$ is maximized at the lattice points and three weaker maxima with same values are located in between nearest-neighbouring lattice points. 
The $E_{\rm eff}$ at the lattice point e.g., $(N_A,N_B)=(1,1)$ can be analytically checked. In addition, let us consider special junctions and gates that are all identical, i.e., $E_{A}=E_{B}=E_s$ and $N_{A}=N_{B}=N_{C}=N_{g}$. Compared with the single-junction case (i.e., $m=1$) with $E_{\mathrm{eff}}$ reduced to the constant $E_{s}$, tuning the reduced offset charge $N_{g}$ towards an integer $k\in \mathbb{N}$ leads to $\lim_{N_g\rightarrow k\in \mathbb{N}} E_{\mathrm{eff}}\left(N_{g}\right)=mE_{s}$, indicating a $m$-fold enhancement of the phase-slip rate.

\subsection{Asymmetry induced by non-integer values of $N_C$}

To investigate how the symmetry breaking affects the coherent quantum phase slips, we first consider the asymmetry induced by a gate voltage $C_{C}$ and capacitance $\nu_{C}$ that leads to non-integer values of $N_{C}(=C_{C}\nu_{C}/2e)$. In Figs.~\ref{fig:alpha5_2_EB1_asymm}(a)-\ref{fig:alpha5_2_EB1_asymm}(c), the effective phase-slip rate from Eq.~(\ref{eq:effective_tunneling}) is presented for three different values of $N_{C}$. It is shown that square lattice patterns similar to Fig.~\ref{fig:alpha5_2_EB1_symm}(a) are generated. 
However, the consideration of a non-integer value of $N_C$, e.g., $0.2$ [Fig.~\ref{fig:alpha5_2_EB1_asymm}(a)] or $0.8$ [Fig.~\ref{fig:alpha5_2_EB1_asymm}(c)] leads to the positions for the maximized $E_{\rm eff}$ slightly shifted away from the lattice points and the three weaker maxima are not identical any more. In particular, we find that the half-integer value, i.e., $N_C=0.5$ considered in Fig.~\ref{fig:alpha5_2_EB1_asymm}(b) gives rise to a vanishing $E_{\rm eff}$ at the lattice points with two splitted maxima around, implying an asymmetry-induced reduction of $E_{\rm eff}$ compared with that in Fig.~\ref{fig:alpha5_2_EB1_symm}(a) for an integer value of $N_C$. 

The main features presented in Figs.~\ref{fig:alpha5_2_EB1_asymm}(a)-\ref{fig:alpha5_2_EB1_asymm}(c) are understandable. Let us consider $E_{\rm eff}$ at $(N_{A}, N_{B})=(0,0)$ as an exanple. Now $\bar{q}=\frac{1}{2}N_{C}$ and $\eta_A=\eta_B=\alpha E_{A}$ [see Eqs.~(\ref{eq:eta_X})-(\ref{eq:bar_q})], so Eq.~(\ref{eq:effective_tunneling}) gives $E_{\rm eff}/E_A=2\alpha|\cos(\pi N_{C})|$ which leads to $8.09$, $0$ and $8.09$ when $\alpha=5$ and $N_{C}=0.2$, $0.5$ and $0.8$, respectively, as shown in Figs.~\ref{fig:alpha5_2_EB1_asymm}(a)-\ref{fig:alpha5_2_EB1_asymm}(c). Moreover, it equals $10$ for $N_{C}=1$ in Fig.~\ref{fig:alpha5_2_EB1_symm}(a). 
Due to the periodic dependence of $E_{\rm eff}$ on $N_A$ and $N_B$, this analysis based on the point $(N_A,N_B)=(0,0)$ is generalizable to all lattice points. 
We additionally notice that there is a mirror symmetry between Figs.~\ref{fig:alpha5_2_EB1_asymm}(a) and \ref{fig:alpha5_2_EB1_asymm}(c) with respect to the white dashed line corresponding to $N_A+N_B=0$. This symmetry results from the facts that Eq.~(\ref{eq:bar_q}) reduces to $\bar{q}_i=N_{Ci}/2+N_{Bi}+N_{Ai}$ where $i=a,c$ corresponding to Figs.~\ref{fig:alpha5_2_EB1_asymm}(a) and \ref{fig:alpha5_2_EB1_asymm}(c), respectively, and assuming the mirror symmetry together with $\cos(4\pi\bar{q}_i)$ in $E_{\rm eff}$ [i.e., Eq.~(\ref{eq:effective_tunneling})] implies $N_{Aa}+N_{Ba}=-N_{Ac}-N_{Bc}$, $\bar{q}_c+\bar{q}_a=k/2$, and in particular $N_{Cc}+N_{Ca}=k$ with $k$ being an integer. This condition holds for $N_C=N_{Ca}=0.2$ [Fig.~\ref{fig:alpha5_2_EB1_asymm}(a)] and $N_C=N_{Cc}=0.8$ [Fig.~\ref{fig:alpha5_2_EB1_asymm}(c)].

\subsection{Asymmetric numbers of junctions}

In this subsection, we further consider the asymmetry induced by unequal numbers in the two sets of phase-slip junctions. 
Similar to Figs.~\ref{fig:alpha5_2_EB1_symm}(a) and \ref{fig:alpha5_2_EB1_asymm}(a)-\ref{fig:alpha5_2_EB1_asymm}(c) for the symmetric and asymmetric setups induced by an integer and non-integer values of $N_C$ respectively, Figs.~\ref{fig:alpha5_2_EB1_symm}(b) and \ref{fig:alpha5_2_EB1_asymm}(d)-\ref{fig:alpha5_2_EB1_asymm}(f) show the phase-slip rate $E_{\rm eff}$ for the asymmetric setup with $\alpha=2$ for totally $m=10$ junctions. This asymmetric configuration drastically changes the behaviors of $E_{\rm eff}$, as demonstrated by stripe patterns in sharp contrast to square lattice patterns in Figs.~\ref{fig:alpha5_2_EB1_symm}(a) and \ref{fig:alpha5_2_EB1_asymm}(a)-\ref{fig:alpha5_2_EB1_asymm}(c). In particular, instead of a vanishing phase slip with $E_{\rm eff}=0$ at $N_{A}=N_{B}=0$ and $N_{C}=0.5$ for $\alpha=5$ in Fig.~\ref{fig:alpha5_2_EB1_asymm}(b), Fig.~\ref{fig:alpha5_2_EB1_asymm}(e) shows finite values of the phase-slip rate $E_{\rm eff}$ induced by the asymmetry (i.e., $\alpha\neq m/2$). This does show the asymmetric number of junctions can increase collective phase slips. 
In addition, similar to the mirror symmetry between patterns of Figs.~\ref{fig:alpha5_2_EB1_asymm}(a) and \ref{fig:alpha5_2_EB1_asymm}(c), Figs.~\ref{fig:alpha5_2_EB1_asymm}(d) and \ref{fig:alpha5_2_EB1_asymm}(f) are also symmetric to each other with respect to the point $(N_A, N_B)=(0,0)$. 
This symmetry requires $N_{Ad}+N_{Af}=N_{Bd}+N_{Bf}=0$ and $\cos(4\pi\bar{q}_d)=\cos(2k\pi-4\pi\bar{q}_f)$ where $k$ is an integer and $\bar{q}_i=N_{Ci}/2+7N_{Bi}/4+N_{Ai}/4$ with $i=d,f$ corresponding to Figs.~\ref{fig:alpha5_2_EB1_asymm}(d) and \ref{fig:alpha5_2_EB1_asymm}(f), respectively. Then, we have $\bar{q}_d+\bar{q}_f=(N_{Cd}+N_{Cf})/2=k/2$ which holds for Figs.~\ref{fig:alpha5_2_EB1_asymm}(d) and \ref{fig:alpha5_2_EB1_asymm}(f).

\begin{figure}
\includegraphics[width=0.85\columnwidth]{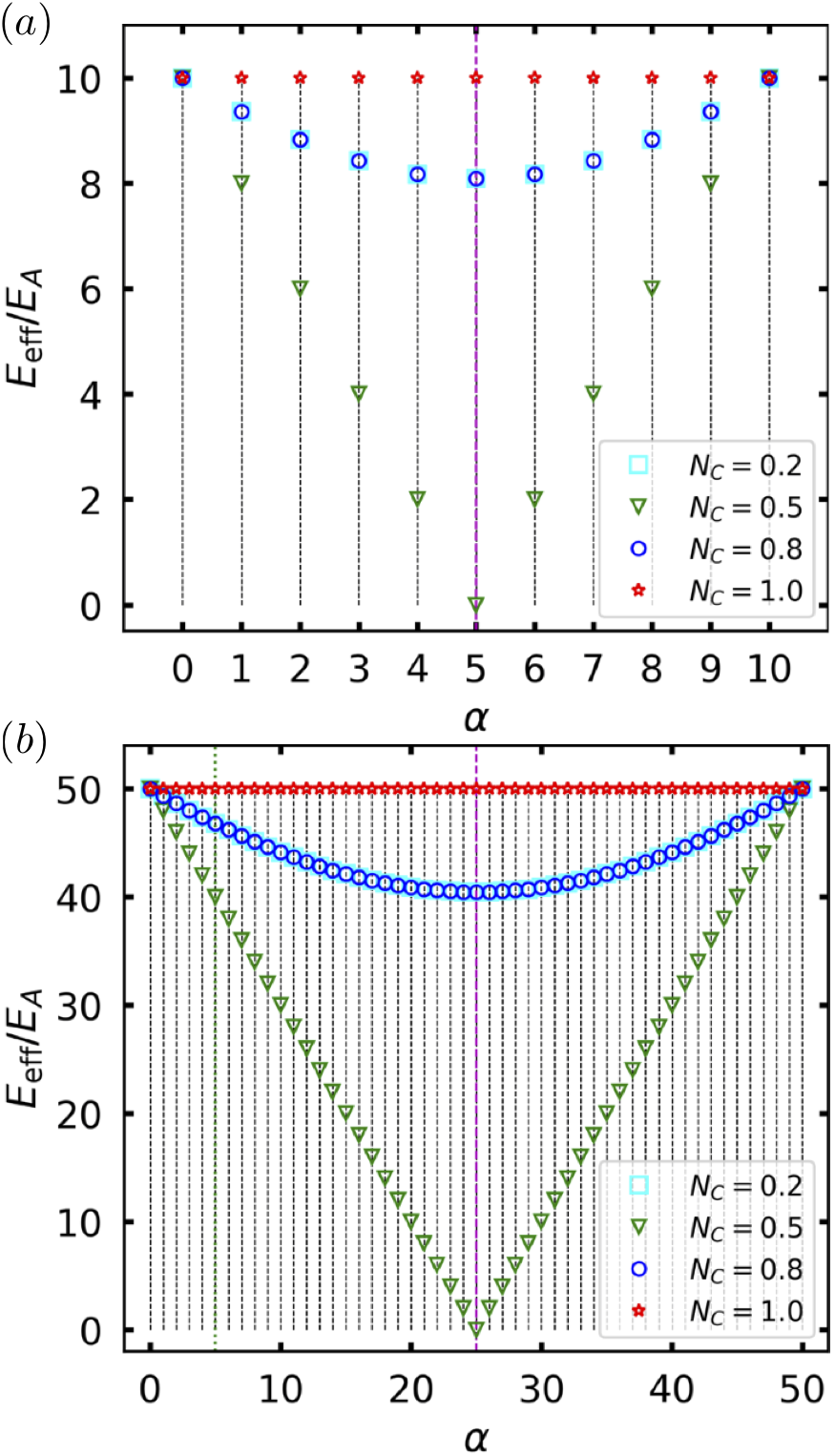}
\caption{The effective phase-slip rate $E_{\rm eff}$ as a function of $\alpha$ for a given total number of junctions (a) $m=10$ and (b) $m=50$ with $N_A=N_B=1.0$ and $E_A=E_B$. Other parameters are same as in Fig.~\ref{fig:alpha5_2_EB1_symm}(a). The vertical dashed line in magenta indicates the position of symmetric number of junctions, i.e., $\alpha/m=1/2$, and the degree of asymmetry becomes maximal when $\alpha\rightarrow0$ or $m$ while it is minimal for $\alpha=m/2$ for $m$ being even.} 
\label{fig:alpha_EB1}
\end{figure}

In order to have a better understanding of the asymmetry effect due to unequal numbers of two types of nanowire junctions, in Fig.~\ref{fig:alpha_EB1} we demonstrate how the effective phase-slip rate changes when varying $\alpha/m$. 
To be consistent with the consideration above, let us first consider a total number of junctions $m=10$ in Fig.~\ref{fig:alpha_EB1}(a). Figure~\ref{fig:alpha_EB1}(a) shows, for $N_C=1.0$, an independence of $E_{\rm eff}$ on $\alpha$, implying that an asymmetric setup ($\alpha\neq m/2$) is as good as a symmetric one ($\alpha=m/2$), both of which gives $10$-fold enhancement. For $N_C=0.2,0.5,0.8$, $E_{\rm eff}$ is, however, minimized at $\alpha=5$ [see the vertical dashed line in magenta in Fig.~\ref{fig:alpha_EB1}(a)]. The further increase of the degree of asymmetry (i.e., far away from $\alpha=m/2$) for a fixed $N_C$ increases $E_{\rm eff}$, indicating a better performance of an asymmetric setup than that of a symmetric one, although it is limited by the $m$-fold enhancement, i.e., $E_{\rm eff}/E_A \le m$ for $E_A=E_B$ and $N_A=N_B=1.0$. Note that these results at $\alpha=5$ and $2$ are consistent with observations at the lattice point $(N_A,N_B)=(1.0,1.0)$ in Figs.~\ref{fig:alpha5_2_EB1_symm}(a), \ref{fig:alpha5_2_EB1_asymm}(a)-\ref{fig:alpha5_2_EB1_asymm}(c), and \ref{fig:alpha5_2_EB1_symm}(b), \ref{fig:alpha5_2_EB1_asymm}(d)-\ref{fig:alpha5_2_EB1_asymm}(f), respectively. 

Since the degree of an asymmetry (e.g., $|\alpha/m-1/2|$) can increase by using a large $m$, we consider $m=50$ in Fig.~\ref{fig:alpha_EB1}(b).  
Similar to Fig.~\ref{fig:alpha_EB1}(a), a symmetric setup, i.e., $\alpha=m/2$ (indicated by the vertical dashed line in magenta) shows a minimized $E_{\rm eff}$ for a given $N_C$ and the increase of an asymmetry away from $\alpha=25$ (so that $\alpha/m=0.5$) enhances phase slips. 
In addition, for a given asymmetry, e.g., $\alpha/m=1/5$ together with a fixed $N_C=0.5$, Fig.~\ref{fig:alpha_EB1}(b) shows the $30$-fold enhancement at $\alpha=10$ which is larger than the $6$-fold enhancement for $\alpha=2$ in Fig.~\ref{fig:alpha_EB1}(a), indicating an advantage of using multiple junctions. 
Finally, we notice that the $m$-fold enhancement at $\alpha=0$ and $\alpha=m$ in Figs.~\ref{fig:alpha_EB1}(a) and \ref{fig:alpha_EB1}(b) are consistent with our analytical result obtained in Sec.~\ref{sec:symm}. 

\begin{figure}
\includegraphics[width=0.9\columnwidth]{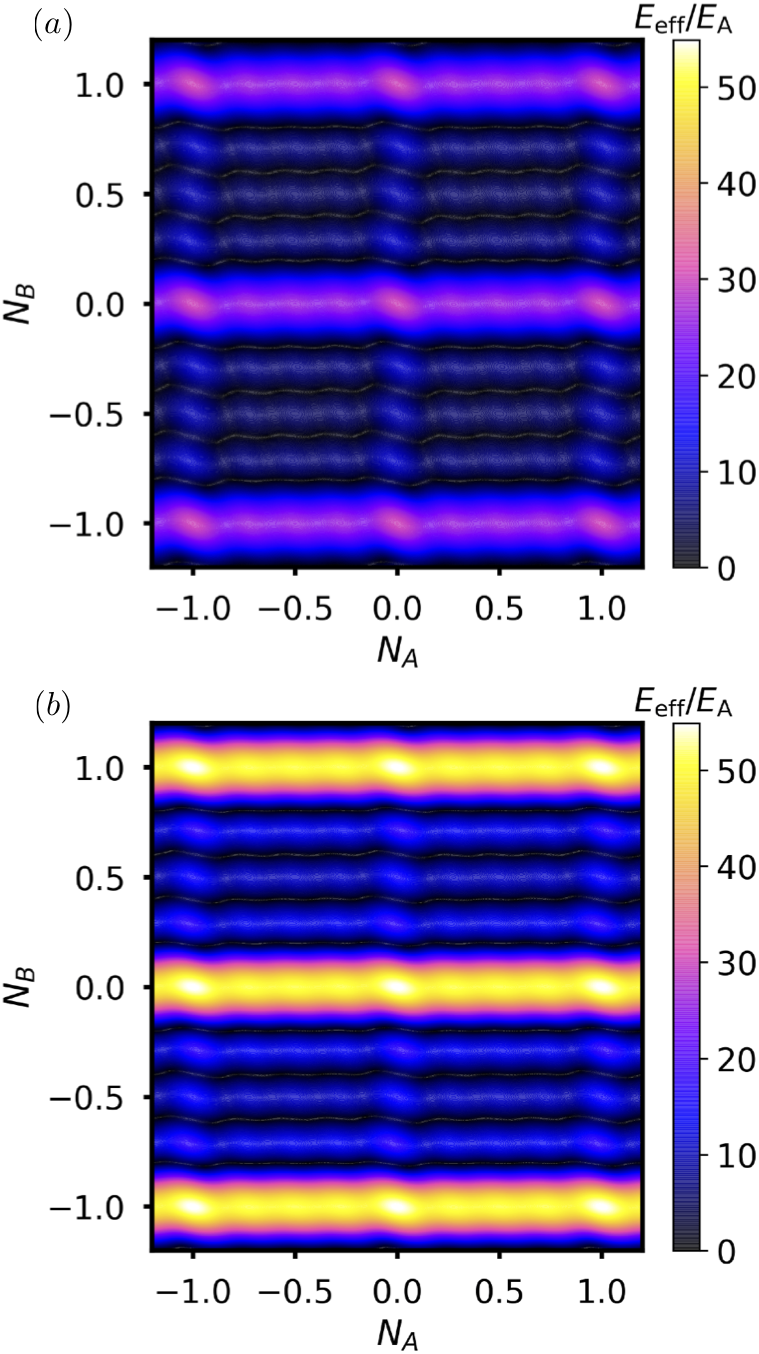}
\caption{The effective phase-slip rate $E_{\mathrm{eff}}$ of an asymmetric multi-junction phase-slip flux qubit (in units of $E_{A}$) versus the reduced offset charges $N_{A}(\equiv C_{A}\nu_{A}/2e)$ and $N_{B}(\equiv C_{B}\nu_{B}/2e)$ for (a) $E_{B}/E_{A}=5$ and (b) $E_{B}/E_{A}=10$ given that $N_{C}=1$, $\alpha=5$, and $m=10$.  Other parameters are same as Fig.~\ref{fig:alpha5_2_EB1_symm}(a). Note that the symmetric case of $E_{B}/E_{A}=1$ has been presented in Fig.~\ref{fig:alpha5_2_EB1_symm}(a).} 
\label{fig:E_S_eff_vs_ES_BA}
\end{figure}

\subsection{Asymmetric junction slip rates}

In Fig.~\ref{fig:E_S_eff_vs_ES_BA}, we show how the asymmetry induced by $E_{B}/E_{A}$ affects the effective phase-slip rate. In contrast to Fig.~\ref{fig:alpha5_2_EB1_symm}(a) with $E_{B}/E_{A}=1$ that shows behaviors in a square lattice for a symmetric setup, 
the considerations of $E_{B}/E_{A}=5$ in Fig.~\ref{fig:E_S_eff_vs_ES_BA}(a) and $E_{B}/E_{A}=10$ in Fig.~\ref{fig:E_S_eff_vs_ES_BA}(b) significantly change the behaviors of $E_{\rm eff}$, and in particular the latter gives much larger $E_{\rm eff}$ at, e.g., the lattice point $(N_A,N_B)=(0,0)$. This asymmetry effect due to $E_{\rm B}/E_{\rm A}\neq 1$ is also quite different from that resulting from $\alpha/m\neq 1/2$ in Figs.~\ref{fig:alpha5_2_EB1_symm}(b) and \ref{fig:alpha5_2_EB1_asymm}(d)-\ref{fig:alpha5_2_EB1_asymm}(f), compared with the fully symmetric case in Fig.~\ref{fig:alpha5_2_EB1_symm}(a) with $E_{B}=E_{A}$ and $\alpha= m/2$. 

\begin{figure}
\includegraphics[width=0.8\columnwidth]{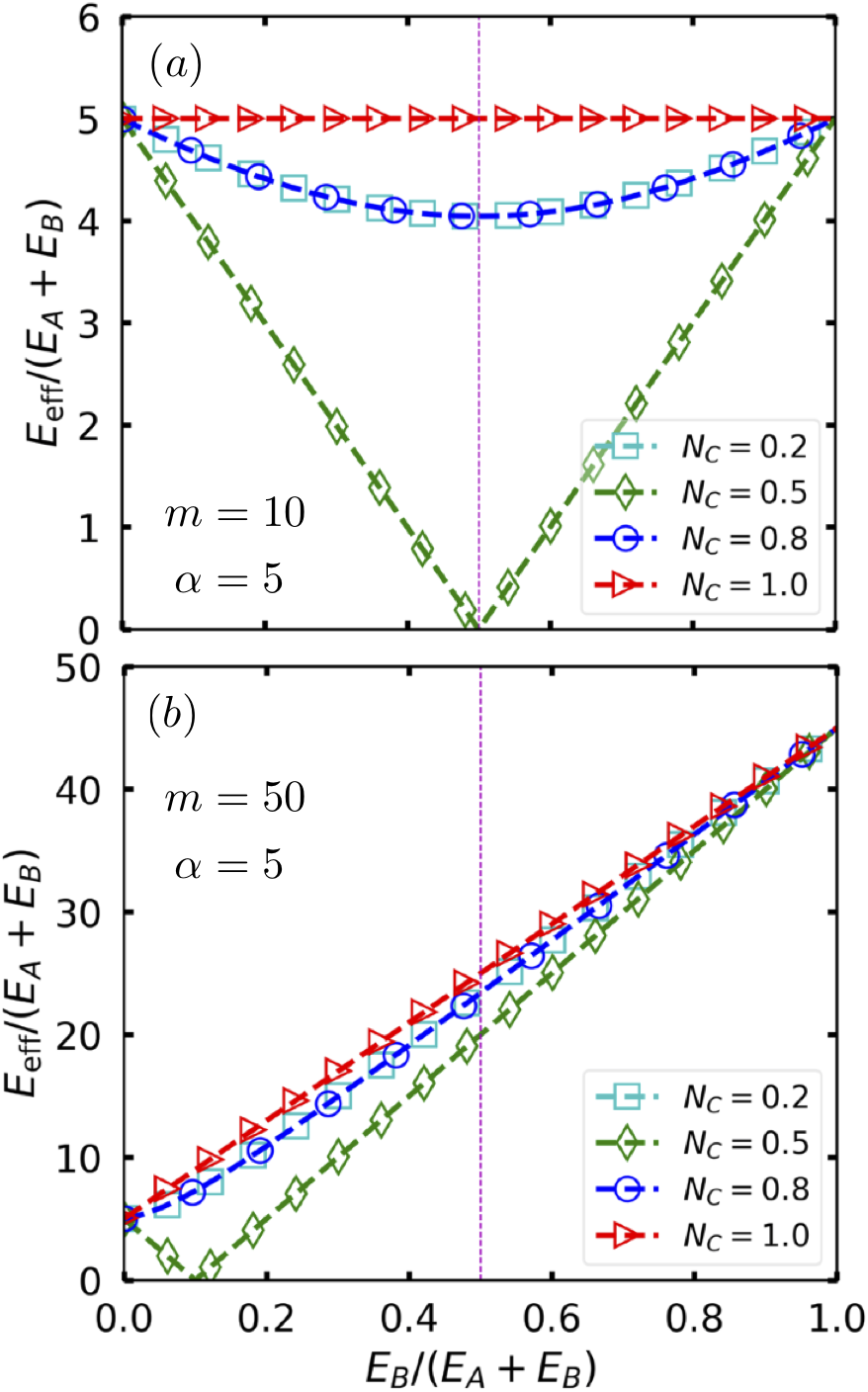}
\caption{The effective phase-slip rate $E_{\mathrm{eff}}/(E_{ A}+E_{ B})$ as a function of the asymmetry characterised by $E_{ B}/(E_{ A}+E_{ B})$ at $N_{A}=N_{B}=1$ for (a) $m=10$ and (b) $m=50$ with $\alpha=5$. The vertical dashed lines in magenta indicate the symmetry positions. Other parameters are same as in Fig.~\ref{fig:alpha5_2_EB1_symm}(a).}
\label{fig:E_S_eff_vs_ESBA}
\end{figure}

Besides two examples of $E_B/E_A$ considered in Fig.~\ref{fig:E_S_eff_vs_ES_BA}, in Fig.~\ref{fig:E_S_eff_vs_ESBA}(a) we particularly consider the response of the effective phase-slip rate to continuously varying the asymmetry characterized by $E_B/(E_A+E_B)$. 
When the gate voltages and capacitances are fixed at e.g., $N_{A}=N_{B}=1$, Figure~\ref{fig:E_S_eff_vs_ESBA}(a) shows that, with the increase of the asymmetry [i.e.,  varying $E_{B}/(E_{A}+E_B)$ away from $0.5$ indicated by the vertical dashed line in magenta], $E_{\rm eff}/(E_A+E_B)$ increases monotonically for $N_{C}=0.2,0.5,0.8$, while for $N_{C}=1.0$, it does not change. 
When considering $E_A=E_B$, these results at the symmetry position [i.e., $E_B/(E_A+E_B)=0.5$] are consistent with those with $\alpha=5$ in Fig.~\ref{fig:alpha_EB1}(a).  
Besides $m=10$, we additionally consider the total number of junctions $m=50$ [just like that in Fig.~\ref{fig:alpha_EB1}(b)] in Fig.~\ref{fig:E_S_eff_vs_ESBA}(b) with the same $\alpha$ as in Fig.~\ref{fig:E_S_eff_vs_ESBA}(a). It is shown that, in contrast to Fig.~\ref{fig:E_S_eff_vs_ESBA}(a), the minimised $E_{\rm eff}/(E_A+E_B)$ does not appear at $E_B/(E_A+E_B)=0.5$. This is due to the symmetry already broken by unequal numbers of junctions i.e., $\alpha/m\neq1/2$ in Fig.~\ref{fig:E_S_eff_vs_ESBA}(b). The comparison between Figs.~\ref{fig:E_S_eff_vs_ESBA}(a) and \ref{fig:E_S_eff_vs_ESBA}(b) additionally shows that the increase of number of total junctions can lead to an enhanced $E_{\rm eff}$ for a given $\alpha=5$. 
In addition, the results at the symmetry position indicated by the vertical dashed lines in magenta in Figs.~\ref{fig:E_S_eff_vs_ESBA}(b) become the ones at $\alpha=5$ (indicated by the dotted line in green) in Fig.~\ref{fig:alpha_EB1}(b), if $E_A=E_B$ is assumed. 
In a word, compared with the results at $E_B/(E_A+E_B)=0.5$, Figure~\ref{fig:E_S_eff_vs_ESBA} shows that much larger $E_{\rm eff}/(E_A+E_B)$ can appear at $E_B/(E_A+E_B)\neq 0.5$, indicating a better performance of an asymmetric setup than that of a symmetric one is available.

\subsection{Further remarks}

We now briefly discuss about a single phase-slip junction in order to highlight the advantages of using multiple phase-slip junctions. For a single junction, the phase-slip rate $E_{s}$ characterizing coherent quantum phase slips in a phase-slip flux qubit follows~\cite{Lau01PRL,MooijHarmans05NJP}:
\begin{equation}
E_{s}=1.5c\frac{k_{B}T_{c}\zeta}{\hbar \xi }\sqrt{N_{\xi }}e^{-0.3dN_{\xi }},
\label{eq:tunnelingrate}
\end{equation}
where $\zeta$ and $\xi $ are the physical length and the coherence length of the superconducting wire, respectively, and $T_{c}$ is the critical temperature. The constants $c$
and $d$ are of order unity~\cite{MooijHarmans05NJP}. Also, $N_{\xi } \equiv R_{q}/R_{\xi }$ is the number of effective conductive channels (or dimensionless conductance) defined by the ratio between the resistance quantum $R_{q}\equiv h/4e^{2} $ and the resistance $R_{\xi }$ of a superconducting wire. 
In order to increase $E_{s}$, one can raise the factor $e^{-0.3d N_{\xi}}\sqrt{N_{\xi}}/\xi$. This requires using a disordered material with a small $N_{\xi}$ for the junction. However, the increase of the phase-slip rate by raising the disorders in the junction is limited because too many disorders in the material can lead to strong Coulomb interactions and thus destroy the superconductivity~\cite{Finkelstein94PhysB}. Alternatively, one can increase the phase-slip rate $E_{s}$ by using a longer junction [see Eq.~(\ref{eq:tunnelingrate})]. However, the enhancement is also limited because the quantum fluctuations needed for the emergence of the phase slips become weakened and can even disappear as the junction becomes long. For example, the length of the junction can only be as long as $\zeta\sim 200$ nm for a $\ce{MoGe}$ nanowire~\cite{Lau01PRL,Bezryadin08JPCM}.

Instead of using a single junction, we can use multiple phase-slip junctions that allows for example asymmetric configurations to achieve a large effective phase-slip rate $E_{\mathrm{eff}}$ under certain parameters according to Eq.~(\ref{eq:effective_tunneling}) as demonstrated above, although each junction may have a small phase-slip rate $E_{s}$. Both the disorders and the length are within the allowed range for the phase slips to occur. Indeed, a single junction may not exhibit an appreciable phase slip when its phase-slip rate is small, but our result reveals that a large phase slip can be achieved by multiple junctions acting collectively. Therefore, the multiple-junction setup not only can achieve a large effective phase-slip
rate to demonstrate appreciable phase slips, but also could enable the use of materials with weak phase slips in superconducting quantum circuits.

\section{Two inductively coupled multi-junction phase-slip flux qubits \label{sec:coupledqubits}}

In order to implement a nontrivial two-qubit quantum gate, one needs a pair of coupled phase-slip flux qubits. 
It has been experimentally demonstrated that two inductively coupled fluxonium atoms constitute a fluxonium-based artificial molecule that shows a tunable magnetic dipole or quadrupole moment~\cite{KouDevoret17PRX}.
In analogue, here we consider two multi-junction phase-slip flux qubits coupled via a mutual inductance as shown in Fig.~\ref{fig:QPSJ_fig}(b).
Adopting again two sets of junction parameters [see Eq.~(\ref{eq:QPSJ_AB})], the phase drop across the $i$th phase-slip junction in the left (right)
superconducting loop is
\begin{eqnarray}
\gamma _{i,1 ( 2) } &=& \Biggr\{ \begin{array}{ll}
\gamma_{A, 1(2)} = 2\pi \frac{L_{kA,1(2)} I_{1(2)}}{\Phi_0} & \text{ if } 1\le i\le\alpha_{1(2)} \\
\gamma_{B, 1(2)} = 2\pi \frac{L_{kB,1(2)} I_{1(2)}}{\Phi_0} & \text{ if } \alpha_{1(2)} < i\le m_{1(2)} 
\end{array}
 \label{eq:kinetic_coupled}
\end{eqnarray}
where $L_{ki,1\left( 2\right) }$ is the kinetic inductance of the $i$th junction in the left (right) phase-slip flux qubit, and $I_{1\left( 2\right) }$ is the corresponding supercurrent. The phase drop related to the geometric inductance of each loop is
\begin{equation}
\gamma _{g,1\left( 2\right) }=2\pi \frac{L_{g,1\left( 2\right) }I_{1\left(
2\right) }}{\Phi _{0}}.  \label{eq:geometry_coupled}
\end{equation}
Also, the magnetic flux in each loop is affected by an adjacent loop through the mutual inductance $M$ between them. The resulting phase drops are given by
\begin{equation}
\gamma _{12}=2\pi \frac{M I_{2}}{\Phi _{0}},~~\gamma _{21}=2\pi \frac{M I_{1}}{\Phi _{0}}.  \label{eq:mutual_coupled}
\end{equation}
Now, the fluxoid quantization condition for each superconducting loop becomes
\begin{eqnarray}
2\pi f_{1( 2) } - \gamma_{t,1(2)} +\gamma _{12( 21) } =2\pi n_{1( 2) },  
\label{eq:quantizationII}
\end{eqnarray}
where $\gamma_{t,1} = \alpha_{1} \gamma_{A,1} + (m_{1}-\alpha_{1}) \gamma_{B,1} +\gamma _{g,1}$ and $\gamma_{t,2} = \alpha_{2} \gamma_{A,2} + (m_{2}-\alpha_{2}) \gamma_{B,2} +\gamma _{g,2}$. Here the eigenvalues of $n_{1\left( 2\right) }\equiv \Phi _{1\left( 2\right) }/\Phi _{0}$ are integers, and $f_{1\left( 2\right) }\equiv \Phi _{1\left( 2\right) ,\mathrm{ext}}/\Phi _{0}$ is the reduced external magnetic flux, with $\Phi_{1\left( 2\right) ,\mathrm{ext}}$ being the externally applied flux in the left (right) loop.

From Eqs.~(\ref{eq:kinetic_coupled})-(\ref{eq:mutual_coupled}), we have two coupled equations 
\begin{eqnarray}
L_{1}I_{1}-M I_{2}&=&\frac{\Phi _{0}}{2\pi }( \gamma_{t,1} - \gamma _{12}) ,  \label{eq:current1} \\
L_{2}I_{2}-M I_{1}&=&\frac{\Phi _{0}}{2\pi }( \gamma_{t,2} - \gamma _{21}) ,  \label{eq:current2}
\end{eqnarray}
where the total inductance of the left and the right loops are respectively given by $L_{1}= \alpha_{1} L_{kA,1} + (m_{1}-\alpha_{1}) L_{kB,1}+L_{g,1}$ and $L_{2}= \alpha_{2} L_{kA,2} + (m_{2}-\alpha_{2}) L_{kB,2}+L_{g,2}$. Using the fluxoid quantization condition in Eq.~(\ref{eq:quantizationII}), we can solve Eqs.~(\ref{eq:current1}) and (\ref{eq:current2}) and obtain the currents in two loops as
\begin{eqnarray}
I_{1}&=&\frac{L_{2}\left( f_{1}-n_{1}\right) +M \left( f_{2}-n_{2}\right)
}{\Lambda _{-}}\Phi _{0},  \label{eq:current_i} \\
I_{2}&=&\frac{L_{1}\left( f_{2}-n_{2}\right) +M \left( f_{1}-n_{1}\right)
}{\Lambda _{-}}\Phi _{0},  \label{eq:current_j}
\end{eqnarray}
where $\Lambda _{-}=L_{1}L_{2}-M^{2}$.

Similar to the single phase-slip flux qubit (see Appendix \ref{sec:Append_qubit}), the Hamiltonian of the two inductively coupled phase-slip flux qubits is given by
\begin{eqnarray}
H_{q-q}&=&\sum_{j=1}^{2}[\frac{1}{2} L_{j}I_{j}^{2}-E_{{\rm eff},j}\cos(\varphi_j) \cos( 2\pi Q_{j}) \notag\\
&&-E_{{\rm eff},j}\sin(\varphi_j) \sin( 2\pi Q_{j}) ] 
+M I_{1}I_{2},
\label{eq:qqHamiltonian}
\end{eqnarray}
with 
\begin{eqnarray}
E_{{\rm eff},j} &=& \sqrt{\eta_{A,j}^2+\eta_{B,j}^2 +2\eta_{A,j}\eta_{B,j}\cos(4\pi\bar{q}_j)}, \label{eq:effective_tunneling2}\\ 
Q_j&=&\frac{1}{2} \left( \sum_{i=1}^{\alpha_j}\frac{q_{i, j}}{\alpha_j} +\sum_{i=\alpha_j+1}^{m_j} \frac{q_{i,j}}{m_j-\alpha_j} \right), \label{eq:Q_j}\\
\varphi_j &=& \arctan\left[\frac{\eta_{A,j} - \eta_{B,j}}{\eta_{A,j} + \eta_{B,j}} \tan(2\pi\bar{q}_j)\right], \label{eq:varphi2}
\end{eqnarray}
where
\begin{eqnarray}
\eta_{A,j} &=& E_{A,j} \frac{\sin(\alpha_j\pi N_{A,j})}{\sin(\pi N_{A,j})}, \\
\eta_{B,j} &=& E_{B,j} \frac{\sin[(m_j-\alpha_j)\pi N_{B,j}]}{\sin(\pi N_{B,j})}, \\ 
\bar{q}_j &=& \frac{1}{2}N_{C,j} + \frac{m_j-\alpha_j-1}{4} N_{B,j} +\frac{\alpha_j-1}{4} N_{A,j},
\end{eqnarray} 
with $N_{A,j}$, $N_{B,j}$, and $N_{C,j}$ being the reduced offset charge in the $j$th multi-junction phase-slip flux qubit, and $\alpha_j$ the number of one set of junctions with $m_j$ being the total number of junctions in the qubit. In Eq.~ (\ref{eq:Q_j}), $q_{i,1(2)}$ is the number of Cooper pairs having tunnelled through the $i$th phase-slip junction in the left (right) qubit.
In the fluxoid representation, the Hamiltonian~(\ref{eq:qqHamiltonian}) can be written as
\begin{eqnarray}
H_{q-q}&=&\sum_{j=1}^{2}\sum_{n_{j}} \Big[
E_{L,j}\left( n_{j}-f_{j}\right) ^{2} |n_j\rangle\langle n_j| \notag\\
&&-\frac{E_{\mathrm{eff},j}}{2} 
( e^{-i\varphi_j} \vert n_{j}+1\rangle \langle n_{j}\vert 
 + e^{i\varphi_j} \vert n_{j}\rangle \langle n_{j}+1\vert) \Big] \notag \\
&&+E_{12} (n_{1}-f_{1}) ( n_{2}-f_{2}) , 
\label{eq:interactionHamiltonian}
\end{eqnarray}
where
\begin{eqnarray}
E_{L,1\left( 2\right) }&=&\frac{\left( 2\Lambda _{+}-\Lambda _{-}\right)
L_{2\left( 1\right) }}{2\Lambda _{-}^{2}}\Phi _{0}^{2}, \\
E_{12}&=&\frac{2\Lambda_{+}+\Lambda_{-}}{\Lambda _{-}^{2}}\sqrt{\frac{\Lambda _{+}-\Lambda _{-}}{2}}\Phi_{0}^{2},
\end{eqnarray}
with $\Lambda _{+}=L_{1}L_{2}+M^{2}$. 
Similar to a single qubit, the phase in the Hamiltonian of two coupled qubits can be eliminated by using the transformation
$|\widetilde{n}_{1(2)}\rangle = e^{-i\varphi_{1(2)} n_{1(2)}} |n_{1(2)}\rangle$
and we have
\begin{eqnarray}
H_{q-q}&=&\sum_{j=1}^{2} \sum_{\widetilde{n}_{j}} \Big[
E_{L,j}\left( n_{j}-f_{j}\right) ^{2} |\widetilde{n}_j\rangle\langle \widetilde{n}_j| \notag\\
&&- \frac{E_{\mathrm{eff},j} }{2} 
( \vert \widetilde{n}_{j}+1 \rangle \langle \widetilde{n}_{j}\vert 
 + \vert \widetilde{n}_{j}\rangle \langle \widetilde{n}_{j}+1\vert) \Big] \notag \\
 &&+E_{12} (\widetilde{n}_{1}-f_{1}) ( \widetilde{n}_{2}-f_{2}) .
\end{eqnarray}
These two inductively coupled phase-slip flux qubits are dual to two capacitively coupled charge qubits~\cite{Pashkin03nature,YouHuNori05prb}.

For a multi-junction phase-slip flux qubit described by Hamiltonian~(\ref{eq:Hamiltonian_2}), if it is in the flux regime with $E_{L}\gg E_{\mathrm{eff}}$, the two fluxoid states $|0\rangle $ and $|1\rangle $ are important when $f$ is tuned to be around the optimal point $f\sim \frac{1}{2}$. The Hamiltonian~(\ref{eq:Hamiltonian_2}) can be reduced to $H=E_{L}\left( f-\frac{1}{2}\right)\sigma _{z}-\frac{1}{2}E_{\mathrm{eff}}\sigma _{x}$, where $\sigma_{z}=|0\rangle \left\langle 0\right\vert -|1\rangle \left\langle 1\right\vert $, and $\sigma _{x}=|0\rangle \left\langle 0\right\vert+|1\rangle \left\langle 1\right\vert $~\cite{MooijHarmans05NJP}. For the two inductively coupled phase-slip flux qubits, let us also consider the flux regime with $E_{L,j}\gg E_{\mathrm{eff},j}, E_{12}$. Around the optimal point $f_{j}\sim \frac{1}{2}$ for each qubit, the Hamiltonian~(\ref{eq:interactionHamiltonian}) is reduced to 
\begin{eqnarray}
H_{q-q} &=&\left[ E_{L,1}\left(f_{1} -\frac{1}{2}\right) -\frac{1}{2}E_{12}\left(
f_{2}-\frac{1}{2}\right) \right] \sigma _{z}^{\left( 1\right) }  \notag \\
&&+\left[ E_{L,2}\left(f_{2} -\frac{1}{2}\right) -\frac{1}{2}E_{12}\left(
f_{1}-\frac{1}{2}\right) \right] \sigma _{z}^{\left( 2\right)}  \notag \\
&&-\frac{1}{2}\sum_{j=1}^{2}E_{\mathrm{eff},j}\sigma _{x}^{\left( j\right) }
+\frac{1}{4}E_{12}\sigma_{z}^{\left(1\right)}\sigma_{z}^{\left(2\right)}.   \label{eq:coupledHamiltonian}
\end{eqnarray}
From Eq.~(\ref{eq:coupledHamiltonian}), it can be seen that the mutual inductance yields a ZZ-type interaction between the two phase-slip flux qubits. Also, it shifts the energy level of each qubit.

\section{Discussions and Conclusions \label{sec:conclusions}}

The electrostatic gates that enable one to tune the charges on the superconducting islands in our proposed device unavoidably cause cross-talk, and therefore in practice, it seems difficult for such tuning to be realized very reliably in experiments. One recent experiment on a semiconductor quantum dot array, however, shows that it is possible to eliminate cross-talk through the definition of virtual gates~\cite{Hensgens17nature}. 
In addition, these gates may polarize possible inhomogeneities located randomly inside the phase-slip junctions~\cite{HongistoZorin12PRL,VanevicNazarov12PRL}. This gives rise to ineffective gating and the charge fluctuations on inhomogeneities increase the decoherence of a qubit. However, the recent experimental realization of a charge quantum interference device that contains an island which separates two junctions indicates that the inhomogeneities are possibly not strong enough to destroy the coherence~\cite{GraafAstafiev18nphys}. 

It would be expected that the offset charge fluctuations and charge noise on the islands of phase-slip junctions possibly affect the performance and also experimental realization of our proposed device. 
We notice that, besides enhancing the effective phase-slip rate, the increase of an asymmetry $|\alpha/m-1/2|$ or $E_B/(E_A+E_B)$, as shown in Fig.~\ref{fig:alpha_EB1}(b) or \ref{fig:E_S_eff_vs_ESBA}(b) respectively, decreases the sensitivity of $E_{\rm eff}$ to $N_C$. This implies an asymmetry-induced reduction of the sensitivity to charge noise on the islands and the largest asymmetry gives a minimal sensitivity.  
In general, charge noise exists on all islands such as those with the gate-induced charge $N_A$ or $N_B$ and its reduction might be also possible. There has been previous works on suppressing the sensitivity to charge noise by using a capacitor to shunt the smaller Josephson junction in a flux qubit~\cite{You07PRB} or the Cooper-pair box in a transmon~\cite{Koch07PRA}. In addition, one can reduce the charge fluctuations by shunting a junction with an array of Josephson junctions that behave effectively like a large inductance~\cite{Manucharyan09science,Koch09PRL,CorleviGuichardHekkingHaviland06PRL}. One would expect that shunting capacitively or inductively  phase-slip junctions with appropriately chosen parameters may provide a possible way of suppressing charge noise and fluctuations on the islands of the junctions. The detail investigation would be performed in the future. 
Although a reduction of offset charge fluctuations and charge noise on multiple islands seems quite challenging, the experimental observation of collective phase slips in Josephson junction arrays does exclude a significant contribution of background charges~\cite{PopGuichard10nphys}. In particular the demonstration of Aharonov-Casher interference in a system of many Josephson junctions in a recent experiment indicates that charges on the islands of the Josephson chain can be controlled with a sufficient precision and therefore are stable enough to enable the chain's collective behavior to be observed at the time scale needed for the measurement~\cite{Pop12PRB}. 
These examples together with the experimental realization of a charge quantum interference device~\cite{GraafAstafiev18nphys} suggest that the charge fluctuations and charge noise on islands may be sufficiently small for a realization of our proposed multi-junction phase-slip flux qubit.  

Although coherent quantum phase slips were explored in Josephson junction circuits~\cite{PopGuichard10nphys,Haviland10nphys,RastelliHekking13PRB,MarcoHekking15PRB,SvetogorovHekking18PRB} and the use of multiple Josephson junctions to improve phase slips was also investigated~\cite{PopGuichard10nphys,Pop12PRB,Manucharyan12PRB}, it has been shown that realizing a quantum current standard via Josephson junction arrays is quite challenging~\cite{Cedergren17PRL}. 
Our study of symmetry and asymmetry effects on collective coherent quantum phase slips is complementary to previous works based on nanowire junctions ~\cite{MooijHarmans05NJP,MooijNazarov06Nphys,AstafievTsai12nature,Peltonen13prb,Peltonen16PRB,GraafAstafiev18nphys,Zhao13CPL}. 
Future work would include an extension of the proposed device to consider for example microwave or spatial modulation~\cite{MarcoHekking15PRB,SvetogorovHekking18PRB} which also helps study the phase-charge duality in one-dimensional superconducting nanowires out of equilibrium~\cite{MooijNazarov06Nphys,Kerman2013NJP,CorleviGuichardHekkingHaviland06PRL,GuichardHekking10PRB}.

We want to emphasize that our device made of multiple nanowire junctions is certainly a nontrivial extension of the previously considered single junction for a phase-slip flux qubit~\cite{MooijHarmans05NJP,MooijNazarov06Nphys} or double junctions used for charge quantum interference device~\cite{GraafAstafiev18nphys,Zhao13CPL}. In sharp contrast to these previous works, our consideration of multiple junctions allows the demonstration of highly nontrivial effects due to the asymmetric configurations as presented in our work. The future work would include an extension to more than two types of phase-slip junctions that would certainly lead to rich and interesting results. 
Although effects due to an asymmetry could also be possible in a device made of two junctions, the significantly different result in Fig.~\ref{fig:alpha5_2_EB1_symm}(b) or Figs.\ref{fig:alpha5_2_EB1_asymm}(d)-\ref{fig:alpha5_2_EB1_asymm}(f) and in particular an enhanced effective phase-slip rate induced by varying $\alpha/m$ presented in Fig.~\ref{fig:alpha_EB1}(b), compared with that for $\alpha/m=1/2$ in Fig.~\ref{fig:alpha5_2_EB1_symm}(a), certainly suggests that the multiple junctions with $m>2$ can give rise to more nontrivial results in contrast to those for $m=2$. 

To conclude, we have proposed a new device for studying how quantum phase slips in a phase-slip flux qubit are influenced by the symmetry in multiple nanowire junctions. Our results show that the collective effect of the multiple junctions gives rise to a large phase-slip rate that can lead to an appreciable number of quantum phase slips events. The effective phase-slip rate can be adjusted via the gate voltage on each island between a pair of adjoining phase-slip junctions. Consequently, the phase-slip flux qubit can be controlled by the gate voltages, apart from the magnetic flux applied to the qubit loop. 
Furthermore, we have proposed to couple two multi-junction phase-slip flux qubits via the mutual inductance between them, which are dual to two capacitively coupled charge qubits. Currently, many materials exhibit only weak signals of quantum phase slips, which makes them unsuitable for quantum information processing. Our proposed multi-junction structure not only provides a large effective phase-slip rate that can enhance appreciable signals of quantum phase slips, but also potentially allows those materials to be used as robust elements in superconducting circuits.

\acknowledgements

This work is supported by the National Key Research and Development Program of China (Grant No. 2016YFA0301200), the National Natural Science Foundation of China (Grant Nos. U1801661 and 11774022), and the Hong Kong GRF Grant No. 15301717. T.F.L. is partially supported by Science Challenge Project (Grant No. TZ2018003).

\appendix

\section{Derivation of the Hamiltonian for a multi-junction phase-slip flux
qubit \label{sec:Append_qubit}}

For the multi-junction phase-slip flux qubit, the kinetic energy and the potential energy can be written respectively as
\begin{eqnarray}
T&=&\frac{1}{2} ( L_{g}+\sum\limits_{i=1}^{m}L_{ki} ) I^{2}, \label{eq:kineticE}  \\
U&=&\sum\limits_{i=1}^{m}\int IV_{i}dt=\sum_{i=1}^{m}E_{i} [1-\cos \left( 2\pi q_{i}\right) ],  \label{eq:potentialE} 
\end{eqnarray}
where $L_{g}$ is the geometric inductance of the loop,  $L_{ki}$ is the kinetic inductance, $E_i$ is the phase-slip rate, and $I$, $V_{i}$, $q_i$ denote the supercurrent, the voltage drop, the number of Cooper pairs through the $i$th phase-slip junction, respectively.  
Here we consider two sets of phase-slip junctions that are defined in Eq.~(\ref{eq:QPSJ_AB}). The kinetic energy in Eq.~(\ref{eq:kineticE}) then becomes
\begin{eqnarray}
T&=&\frac{1}{2} [ L_{g}+\alpha L_{kA} +(m-\alpha) L_{kB} ] I^{2},
\end{eqnarray}
and the potential energy can be written as 
\begin{eqnarray}
U&=&U_A+U_B ,
\end{eqnarray}
with 
\begin{eqnarray}
U_A&=& \alpha E_{A} -E_{A} \sum_{i=1}^{\alpha} \cos(2\pi q_i), \label{eq:U_A}\\
U_B&=& (m-\alpha) E_{B} -E_{B} \sum_{i'=\alpha+1}^m \cos(2\pi q_{i'}). \label{eq:U_B}
\end{eqnarray}

Due to the consideration of three sets of values for the capacitance and the gate voltage as defined in Eq.~(\ref{eq:capacitanceVoltage}), the terms in  Eq.~(\ref{eq:U_A}) can be summed to yield the following analytic form:
\begin{eqnarray}
U_A &=& -\frac{E_{A}}{2\sin(2\pi N_{A})} \sum_{i=1}^{\alpha} \{\sin[2\pi(q_i+N_{A})] \notag\\
&&- \sin[2\pi(q_i-N_{A})] \} \notag \\
&=&-\frac{E_{A}}{2\sin(2\pi N_{A})} \{ \sin(2\pi q_{\alpha})+\sin[2\pi(q_{\alpha}+N_{A})] \notag \\
&& -\sin[2\pi(q_1-N_{A})] -\sin(2\pi q_1) \} ,
\label{eq:U_A_1}
\end{eqnarray}
where we have used the relation $q_{i+1}-q_i=N_A (1\le i\le \alpha-1)$ in Eq.~(\ref{eq:chargeImblance}). In order to further simplify this expression, we define the average number of Cooper pairs in the first set of phase-slip junctions
\begin{eqnarray}
q_A&\equiv& \frac{1}{\alpha} \sum_{i=1}^{\alpha} q_{i},
\label{eq:q_A}
\end{eqnarray}
which together with $N_A$ gives rise to two relations
\begin{eqnarray}
q_1 &=& q_A -\frac{\alpha-1}{2} N_A, \\
q_{\alpha} &=& q_A+\frac{\alpha-1}{2} N_A. \label{eq:q_alpha}
\end{eqnarray}
We then substitute $q_1$ and $q_{\alpha}$ in Eq.~(\ref{eq:U_A_1}) with these two expressions and obtain $U_A$ as
\begin{eqnarray}
U_A&=&-\frac{E_{A} \cos(2\pi q_A)}{\sin(2\pi N_{A})} \{ \sin[(\alpha-1)\pi N_{A}] \notag\\
&&+\sin[(\alpha+1)\pi N_{A}] \}\notag\\
&=&-\frac{2 E_{A}\cos(\pi N_{A})}{\sin(2\pi N_{A})} \sin(\alpha\pi N_{A})\cos(2\pi q_A) \notag\\
&=& -\frac{E_{A}\sin(\alpha\pi N_{A})}{\sin(\pi N_{A})} \cos(2\pi q_A). 
\end{eqnarray}
Similarly, $U_B$ in Eq.~(\ref{eq:U_B}) becomes
\begin{eqnarray}
U_B &=&-\frac{E_{B}\sin[(m-\alpha)\pi N_{B}]}{\sin(\pi N_{B})} \cos(2\pi q_B),
\end{eqnarray}
where $q_B$ is the average number of Cooper pairs in the second set of phase-slip junctions defined by
\begin{eqnarray}
q_B &\equiv& \frac{1}{m-\alpha} \sum_{i'=\alpha+1}^{m} q_{i'}, \label{eq:q_B}
\end{eqnarray}
and we have additionally used $q_{i+1}-q_i=N_B (\alpha < i \le m-1)$ in Eq.~(\ref{eq:chargeImblance}), and
\begin{eqnarray}
q_{\alpha+1} &=& q_B - \frac{m-\alpha-1}{2} N_B, \\
q_m &=& q_B + \frac{m-\alpha-1}{2} N_B. \label{eq:q_m}
\end{eqnarray}
We further define two alternative variables, i.e., 
\begin{eqnarray}
Q&\equiv&\frac{q_B + q_A}{2}, \\
\bar{q}&\equiv&\frac{q_B - q_A}{2}. \label{eq:bar_q_app}
\end{eqnarray}
By using Eqs.~(\ref{eq:q_A}), (\ref{eq:q_alpha}), (\ref{eq:q_B}), (\ref{eq:q_m}), and $q_{\alpha+1}-q_{\alpha}=N_{C}$ from Eq.~(\ref{eq:chargeImblance}), $\bar{q}$ in Eq.~(\ref{eq:bar_q_app}) can be evaluated to obtain Eq.~(\ref{eq:bar_q}). 

Thus, the Lagrangian of the multi-junction phase-slip flux qubit is given by
\begin{eqnarray}
L&=&T-U \notag \\
&=&\frac{1}{2} L_t I^{2} + (\eta_A+\eta_B)\cos(2\pi\bar{q}) \cos(2\pi Q) \notag\\
&&+ (\eta_A-\eta_B) \cos(2\pi\bar{q}) \sin(2\pi Q),
\end{eqnarray}
where $L_t=L_g+\alpha L_{kA} +(m-\alpha)L_{kB}$ and $I=2e\dot{q}_{i}=2e\dot{Q}$, because $\dot{q_{i}}=\dot{q}_{i+1}$. Here, $\eta_X$ and $\eta_Y$ are given by Eqs.~(\ref{eq:eta_X}) and (\ref{eq:eta_Y}) respectively. 
We then choose $Q$ as the canonical coordinate. The corresponding canonical momentum is given by 
\begin{eqnarray}
P&=&\frac{\partial L}{\partial \dot{Q}}=2L_t eI.
\end{eqnarray}
By using the fluxoid quantization condition 
\begin{eqnarray}
\Phi_{\rm ext} - L_{t}I &=& n\Phi_0,
\end{eqnarray}
with the external magnetic flux $\Phi_{\rm ext}=f \Phi _{0}$ and the flux due to both geometric and kinetic inductance in the loop $L_{t}I = \gamma _{t} \Phi _{0}/2\pi$, which lead to an equivalent expression $2\pi f-\gamma _{t} = 2\pi n$, 
the supercurrent can 
be expressed as
\begin{eqnarray}
I&=&\frac{\Phi _{0}}{L_t}\left( f-n\right) .
\end{eqnarray}
Therefore, the Hamiltonian of the multi-junction phase-slip flux qubit is obtained as
\begin{eqnarray}
H&=&P\dot{Q}-L \notag\\\
&=&E_{L}\left( n-f\right) ^{2}- (\eta_X+\eta_Y)\cos(2\pi\bar{q}) \cos(2\pi Q) \notag\\
&&- (\eta_X-\eta_Y) \cos(2\pi\bar{q}) \sin(2\pi Q) \notag\\
&=&E_{L}\left( n-f\right) ^{2}- E_{\rm eff} \cos(\varphi) \cos(2\pi Q) \notag\\
&&-E_{\rm eff} \sin(\varphi) \sin(2\pi Q),
\end{eqnarray}
with $E_{L}$, $E_{\rm eff}$, and $\varphi$ given by Eqs.~(\ref{eq:inductiveE}), (\ref{eq:effective_tunneling}), and (\ref{eq:varphi}) respectively. This Hamiltonian gives Eq.~(\ref{eq:Hamiltonian_1}) when expressed in the fluxoid representation.


\begin{thebibliography}{99}
\bibitem{DErrico17PhilTransRSocA} C. D'Errico, S. S. Abbate, G. Modugno, {\it Quantum phase slips: From condensed matter to ultracold quantum gases}, \href{https://doi.org/10.1098/rsta.2016.0425}{Phil. Trans. R. Soc. A {\bf 375}, 20160425 (2017)}. 

\bibitem{Tinkham96} M. Tinkham, \emph{Introduction to Superconductivity}, 2nd ed. (McGraw-Hill, New York, 1996).

\bibitem{Bezryadin13} A. Bezryadin, \emph{Superconductivity in Nanowires}
(Wiley-VCH, 2013).

\bibitem{ArutyunovGolubevZaikin08PhysRep} K. Y. Arutyunov, D. S. Golubev, and A. D. Zaikin, {\it Superconductivity in one dimension}, \href{https://doi.org/10.1016/j.physrep.2008.04.009}{Phys. Rep. \textbf{464}, 1 (2008)}.

\bibitem{NewbowerTinkham1972PRB} R. S. Newbower, M. R. Beasley, and M. Tinkham, {\it Fluctuation effects on the superconductlng transition of tin whisker crystals}, \href{https://doi.org/10.1103/PhysRevB.5.864}{Phys. Rev. B \textbf{5}, 864 (1972)}.

\bibitem{Lau01PRL} C. N. Lau, N. Markovic, M. Bockrath, A. Bezryadin, and M. Tinkham, {\it Quantum phase slips in superconducting nanowires}, \href{https://doi.org/10.1103/PhysRevLett.87.217003}{Phys. Rev. Lett. \textbf{87}, 217003 (2001)}.

\bibitem{Bollinger08PRL} A. T. Bollinger, R. C. Dinsmore III, A. Rogachev, and A. Bezryadin, {\it Determination of the superconductor-insulator phase diagram for one-dimensional wires}, \href{https://doi.org/10.1103/PhysRevLett.101.227003}{Phys. Rev. Lett. \textbf{101}, 227003 (2008)}.

\bibitem{LangerAmbegaokar67PR} J. S. Langer and V. Ambegaokar, {\it Intrinsic resistive transition in narrow superconducting channels}, \href{https://doi.org/10.1103/PhysRev.164.498}{Phys. Rev. \textbf{164}, 498 (1967)}.

\bibitem{McCumberHalperin70PRB} D. E. McCumber and B. I. Halperin, {\it Time scale of intrinsic resistive fluctuations in thin superconducting wires}, \href{https://doi.org/10.1103/PhysRevB.1.1054}{Phys. Rev. B \textbf{1}, 1054 (1970)}.

\bibitem{SkocpolBeasleyTinkham74JLowTempPhys} W. J. Skocpol, M. R. Beasley, and M. Tinkham, {\it Phase-slip centers and nonequilibrium processes in superconducting tin microbridges}, \href{https://doi.org/10.1007/BF00655865}{J. Low. Temp. Phys. \textbf{16}, 145 (1974)}.

\bibitem{MooijHarmans05NJP} J. E. Mooij and C. J. P. M. Harmans, {\it Phase-slip flux qubits}, \href{https://doi.org/10.1088/1367-2630/7/1/219}{New J. Phys. \textbf{7}, 219 (2005)}.

\bibitem{Likharev86} K. K. Likharev, \emph{Dynamics of Josephson Junctions and Circuits} (Gordon and Breach, New York, 1986).

\bibitem{MooijNazarov06Nphys} J. E. Mooij and Y. V. Nazarov, {\it Superconducting nanowires as quantum phase-slip junctions}, \href{https://doi.org/10.1038/nphys234}{Nat. Phys. \textbf{2}, 169 (2006)}.


\bibitem{YouNori11nature} J. Q. You and F. Nori,  {\it Atomic physics and quantum optics using superconducting circuits}, \href{https://doi.org/10.1038/nature10122}{Nature \textbf{474}, 589 (2011)}.

\bibitem{ClarkeWilhelm08nature} J. Clarke and F. K. Wilhelm, {\it Superconducting quantum bits}, \href{https://doi.org/10.1038/nature07128}{Nature \textbf{453}, 1031 (2008)}.

\bibitem{DevoretSchoelkopf13science} M. H. Devoret and R. J. Schoelkopf, {\it Superconducting circuits for quantum information: An outlook}, \href{https://doi.org/10.1126/science.1231930}{Science \textbf{339}, 1169 (2013)}.

\bibitem{KautzLloyd1987APL} R. L. Kautz and F. L. Lloyd, {\it Precision of series-array Josephson voltage standards}, \href{https://doi.org/10.1063/1.98286}{Appl. Phys. Lett. \textbf{51}, 2043 (1987)}, and references therein.

\bibitem{Kerman2013NJP} A. J. Kerman, {\it Flux-charge duality and topological quantum phase fluctuations in quasi-one-dimensional superconductors}, \href{https://doi.org/10.1088/1367-2630/15/10/105017}{New J. Phys. \textbf{15}, 105017 (2013)}.

\bibitem{Zimmerman2005PhysTod} N. M. Zimmerman, {\it Quantum electrical standards}, \href{https://doi.org/10.1063/1.3480089}{Phys. Today \textbf{63}, No.~8, 68 (2010)}.

\bibitem{AstafievTsai12nature} O. V. Astafiev, L. B. Ioffe, S. Kafanov, Y. A. Pashkin, K. Y. Arutyunov, D. Shahar, O. Cohen, and J. S. Tsai, {\it Coherent quantum phase slip}, \href{https://doi.org/10.1038/nature10930}{Nature \textbf{484}, 355 (2012)}.

\bibitem{Bezryadin12nature} A. Bezryadin, {\it Quantum physics: Tunnelling across a nanowire}, \href{https://doi.org/10.1038/484324b}{Nature {\bf 484}, 324 (2012)}.

\bibitem{Peltonen13prb} J. T. Peltonen, O. V. Astafiev, Yu. P. Korneeva, B. M. Voronov, A. A. Korneev, I. M. Charaev, A. V. Semenov, G. N. Golt'sman, L. B. Ioffe, T. M. Klapwijk, and J. S. Tsai, {\it Coherent flux tunneling through NbN nanowires}, \href{https://doi.org/10.1103/PhysRevB.88.220506}{Phys. Rev. B \textbf{88}, 220506 (2013)}.

\bibitem{Peltonen16PRB} J. T. Peltonen, Z. H. Peng, Yu. P. Korneeva, B. M. Voronov, A. A. Korneev, A. V. Semenov, G. N. Gol'tsman, J. S. Tsai, and O. V. Astafiev, {\it Coherent dynamics and decoherence in a superconducting weak link}, \href{https://doi.org/10.1103/PhysRevB.94.180508}{Phys. Rev. B {\bf 94}, 180508 (2016)}. 

\bibitem{HongistoZorin12PRL} T. T. Hongisto and A. B. Zorin, {\it Single-charge transistor based on the charge-phase duality of a superconducting nanowire circuit}, \href{https://doi.org/10.1103/PhysRevLett.108.097001}{Phys. Rev. Lett. \textbf{108}, 097001 (2012)}.

\bibitem{BelkinBezryadin15PRX} A. Belkin, M. Belkin, V. Vakaryuk, S. Khlebnikov, and A. Bezryadin, {\it Formation of quantum phase slip pairs in superconducting nanowires}, \href{https://doi.org/10.1103/PhysRevX.5.021023}{Phys. Rev. X {\bf 5}, 021023 (2015)}. 

\bibitem{Webster13PRB} C. H. Webster, J. C. Fenton, T. T. Hongisto, S. P. Giblin, A. B. Zorin, and P.  A. Warburton, {\it NbSi nanowire quantum phase-slip circuits: dc supercurrent blockade, microwave measurements, and thermal analysis}, \href{https://doi.org/10.1103/PhysRevB.87.144510}{Phys. Rev. B {\bf 87}, 144510 (2013)}. 

\bibitem{Kafanov13JAP} S. Kafanov and N. M. Chtchelkatchev, {\it Single flux transistor: The controllable interplay of coherent quantum phase slip and flux quantization}, \href{https://doi.org/10.1063/1.4818706}{J. Appl. Phys. {\bf 114}, 073907 (2013)}. 

\bibitem{GraafAstafiev18nphys} S. E. de Graaf, S. T. Skacel, T. H\"{o}nigl-Decrinis, R. Shaikhaidarov, H. Rotzinger, S. Linzen, M. Ziegler, U. H\"{u}bner, H. G. Meyer, V. Antonov, E. IlÕichev, A. V. Ustinov, A. Y. Tzalenchuk, and O. V. Astafiev, {\it Charge quantum interference device}, \href{https://doi.org/10.1038/s41567-018-0097-9}{Nat. Phys. {\bf 14}, 590 (2018)}. 

\bibitem{Zhao13CPL} H. Zhao, T. F. Li, J. S. Liu, and W. Chen, {\it Charge-related SQUID and tunable phase-slip flux qubit}, \href{https://doi.org/10.1088/0256-307X/31/3/030303}{Chin. Phys. Lett. {\bf 31}, 030303 (2013)}. 

\bibitem{BellGershenson16PRL} M. T. Bell, W. Zhang, L. B. Ioffe, and M. E. Gershenson, {\it Spectroscopic evidence of the Aharonov-Casher effect in a Cooper pair box}, \href{https://doi.org/10.1103/PhysRevLett.116.107002}{Phys. Rev. Lett. {\bf 116}, 107002 (2016)}. 
\bibitem{ErgulHaviland13NJP} A. Erg\"{u}l, J. Lidmar, J. Johansson, Y. Azizo\v{g}lu, D. Schaeffer, and D. B Haviland, {\it Localizing quantum phase slips in one-dimensional Josephson junction chains}, \href{https://doi.org/10.1088/1367-2630/15/9/095014}{New J. Phys. {\bf 15}, 095014 (2013)}.

\bibitem{PopGuichard10nphys} I. M. Pop, I. Protopopov, F. Lecocq, Z. Peng, B. Pannetier, O. Buisson, and W. Guichard, {\it Measurement of the effect of quantum phase slips in a Josephson junction chain}, \href{https://doi.org/10.1038/nphys1697}{Nat. Phys. {\bf 6}, 589 (2010)}.

\bibitem{Haviland10nphys} D. Haviland, {\it Superconducting circuits: Quantum phase slips}, \href{https://doi.org/10.1038/nphys1747}{Nat. Phys. {\bf 6}, 565 (2010)}. 

\bibitem{Pop12PRB} I. M. Pop, B. Dou\c{c}ot, L. Ioffe, I. Protopopov, F. Lecocq, I. Matei, O. Buisson, and W. Guichard, {\it Experimental demonstration of Aharonov-Casher interference in a Josephson junction circuit}, \href{https://doi.org/10.1103/PhysRevB.85.094503}{Phys. Rev. B {\bf 85}, 094503 (2012)}. 

\bibitem{Manucharyan12PRB} V. E. Manucharyan, N. A. Masluk, A. Kamal, J. Koch, L. I. Glazman, and M. H. Devoret, {\it Evidence for coherent quantum phase slips across a Josephson junction array}, \href{https://doi.org/10.1103/PhysRevB.85.024521}{Phys. Rev. B 85, 024521 (2012)}.

\bibitem{RastelliHekking13PRB} G. Rastelli, I. M. Pop, and F. W. J. Hekking, {\it Quantum phase slips in Josephson junction rings}, \href{https://doi.org/10.1103/PhysRevB.87.174513}{Phys. Rev. B {\bf 87}, 174513 (2013)}. 

\bibitem{MarcoHekking15PRB} A. D. Marco, F. W. J. Hekking, and G. Rastelli, {\it Quantum phase-slip junction under microwave irradiation}, \href{https://doi.org/10.1103/PhysRevB.91.184512}{Phys. Rev. B {\bf 91}, 184512 (2015)}. 

\bibitem{SvetogorovHekking18PRB} A. E. Svetogorov, M. Taguchi, Y. Tokura, D. M. Basko, and F. W. J. Hekking, {\it Theory of coherent quantum phase slips in Josephson junction chains with periodic spatial modulations}, \href{https://doi.org/10.1103/PhysRevB.97.104514}{Phys. Rev. B {\bf 97}, 104514 (2018)}. 

\bibitem{Cedergren17PRL} K. Cedergren, R. Ackroyd, S. Kafanov, N. Vogt, A. Shnirman, and T. Duty, {\it Insulating Josephson junction chains as pinned Luttinger liquids}, \href{https://doi.org/10.1103/PhysRevLett.119.167701}{Phys. Rev. Lett. {\bf 119}, 167701 (2017)}.

\bibitem{Zaikin97prl} A. D. Zaikin, D. S. Golubev, A. van Otterlo, and G. T. Zimanyi, {\it Quantumphase slips and transport in ultrathin superconducting wires}, \href{https://doi.org/10.1103/PhysRevLett.78.1552}{Phys. Rev. Lett. \textbf{78}, 1552 (1997)}.

\bibitem{GolubevZaikin01prb} D. S. Golubev and A. D. Zaikin, {\it Quantum tunneling of the order parameter in superconducting nanowires}, \href{https://doi.org/10.1103/PhysRevB.64.014504}{Phys. Rev. B \textbf{64}, 014504 (2001)}.

\bibitem{Finkelstein94PhysB} A. M. Finkel'stein, {\it Suppression of superconductivity in homogeneously disordered systems}, \href{https://doi.org/10.1016/0921-4526(94)90267-4}{Physica B \textbf{197}, 636 (1994)}.

\bibitem{Bezryadin08JPCM} A. Bezryadin, {\it Quantum suppression of superconductivity in nanowires}, \href{https://doi.org/10.1088/0953-8984/20/04/043202}{J. Phys.: Condens. Matter \textbf{20}, 043202 (2008)}.

\bibitem{Manucharyan09science} V. E. Manucharyan, J. Koch, L. I. Glazman, and M. H. Devoret, {\it Fluxonium: Single Cooper-pair circuit free of charge offsets}, \href{https://doi.org/10.1126/science.1175552}{Science {\bf 326}, 113 (2009)}.

\bibitem{KouDevoret17PRX} A. Kou, W. C. Smith, U. Vool, R. T. Brierley, H. Meier, L. Frunzio, S. M. Girvin, L. I. Glazman, and M. H. Devoret, {\it Fluxonium-based artificial molecule with a tunable magnetic moment}, \href{https://doi.org/10.1103/PhysRevX.7.031037}{Phys. Rev. X {\bf 7}, 031037 (2017)}. 

\bibitem{Pashkin03nature} Y. A. Pashkin, T. Yamamoto, O. Astafiev, Y. Nakamura, D.V. Averin, and J. S. Tsai, {\it Quantum oscillations in two coupled charge qubits}, \href{https://doi.org/10.1038/nature01365}{Nature {\bf 421}, 823 (2003)}.

\bibitem{YouHuNori05prb} J. Q. You, X. Hu, and F. Nori, {\it Correlation-induced suppression of decoherence in capacitively coupled Cooper-pair boxes}, \href{https://doi.org/10.1103/PhysRevB.72.144529}{Phys. Rev. B \textbf{72}, 144529 (2005)}.

\bibitem{Hensgens17nature} T. Hensgens {\it et al}., {\it Quantum simulation of a FermiÐHubbard model using a semiconductor quantum dot array}, \href{https://doi.org/10.1038/nature23022}{Nature {\bf 548}, 70 (2017)}. 

\bibitem{VanevicNazarov12PRL} M. Vanevi\'{c} and Y. V. Nazarov, {\it Quantum phase slips in superconducting wires with weak inhomogeneities}, \href{https://doi.org/10.1103/PhysRevLett.108.187002}{Phys. Rev. Lett. {\bf 108}, 187002 (2012)}. 

\bibitem{You07PRB} J. Q. You, X Hu, S. Ashhab, and F. Nori, {\it Low-decoherence flux qubit}, \href{https://doi.org/10.1103/PhysRevB.75.140515}{Phys. Rev. B {\bf 75}, 140515(R) (2007)}. 

\bibitem{Koch07PRA} J. Koch, T. M. Yu, J. Gambetta, A. A. Houck, D. I. Schuster, J. Majer, A. Blais, M. H. Devoret, S. M. Girvin, and R. J. Schoelkopf, {\it Charge-insensitive qubit design derived from the Cooper pair box}, \href{https://doi.org/10.1103/PhysRevA.76.042319}{Phys. Rev. A {\bf 76}, 042319 (2007)}.

\bibitem{Koch09PRL} J. Koch, V. Manucharyan, M. H. Devoret, and L. I. Glazman, {\it Charging effects in the inductively shunted Josephson junction}, \href{https://doi.org/10.1103/PhysRevLett.103.217004}{Phys. Rev. Lett. {\bf 103}, 217004 (2009)}. 

\bibitem{CorleviGuichardHekkingHaviland06PRL} S. Corlevi, W. Guichard, F. W. J. Hekking, and D. B. Haviland, {\it Phase-charge duality of a Josephson junction in a fluctuating electromagnetic environment}, \href{https://doi.org/10.1103/PhysRevLett.97.096802}{Phys. Rev. Lett. {\bf 97}, 096802 (2006)}. 

\bibitem{GuichardHekking10PRB} W. Guichard and F. W. J. Hekking, {\it Phase-charge duality in Josephson junction circuits: Role of inertia and effect of microwave irradiation}, \href{https://doi.org/10.1103/PhysRevB.81.064508}{Phys. Rev. B 81, 064508 (2010)}.

\end{thebibliography}
\end{document}